\documentclass[a4paper,11pt]{article}
\pdfoutput=1
\usepackage{jcappub}
\usepackage[T1]{fontenc}

\usepackage{epstopdf}

\usepackage{graphicx}
\DeclareGraphicsExtensions{.eps}

\title{\boldmath Caustics of $1/r^n$ binary gravitational lenses: from galactic haloes to exotic matter}

\author[a,b]{V. Bozza}
\author[a,b]{C. Melchiorre}

\affiliation[a]{Dipartimento di Fisica ``E.R. Caianiello'', Universit\`{a} di Salerno,
Via Giovanni Paolo II 132, I-84084 Fisciano (SA), Italy.}
\affiliation[b]{Istituto Nazionale di Fisica Nucleare, Sezione di Napoli, Italy.}

\emailAdd{valboz@physics.unisa.it}
\emailAdd{cmelchiorre@unisa.it}

\abstract{We investigate the caustic topologies for binary gravitational lenses made up of two objects whose gravitational potential declines as $1/r^n$. With $n<1$ this corresponds to power-law dust distributions like the singular isothermal sphere. The $n>1$ regime can be obtained with some violations of the energy conditions, one famous example being the Ellis wormhole. Gravitational lensing provides a natural arena to distinguish and identify such exotic objects in our Universe. We find that there are still three topologies for caustics as in the standard Schwarzschild binary lens, with the main novelty coming from the secondary caustics of the close topology, which become huge at higher $n$. After drawing caustics by numerical methods, we derive a large amount of analytical formulae in all limits that are useful to provide deeper insight in the mathematics of the problem. Our study is useful to better understand the phenomenology of galaxy lensing in clusters as well as the distinct signatures of exotic matter in complex systems.}

\begin{document}
\maketitle
\flushbottom

\section{Introduction}
\label{sec:intro}
Gravitational lensing is certainly one of the most studied phenomena in the physics literature. As light is the main carrier of astrophysical information, its deflection by gravitational fields provides a unique opportunity to measure the mass of astronomical objects and probe the distribution of matter without relying on its emission. There are famous examples in which gravitational lensing has demonstrated the existence of dark matter concentrations in clusters of galaxies \cite{Dark1,Dark2,Dark3} and it is currently used to understand the formation of structures in the Universe \cite{Struct1,Struct2}. Thanks to its ability to unveil the invisible, gravitational lensing has also been invoked to probe the existence of non-ordinary matter in the form of isolated objects with peculiar metrics. A broad literature exists on lensing by any kind of non-Schwarzschild metrics \cite{Zak,BozzaBH}. While these metrics predict strong deviation from Schwarzschild in the strong deflection regime, they typically reduce to Schwarzschild in the weak field limit, leaving only subdominant tiny corrections in the most explored regime. However, there is a considerable exception: the Ellis wormhole \cite{Ellis}. This metric, introduced so as to demonstrate the possibility to connect two asymptotically flat regions by a non-singular throat, has the interesting property of falling down as $1/r^2$ asymptotically. Thus, any observables depending on the gravitational field is affected, even in the weak field limit. Concentrating on the gravitational lensing effect, it has been proved that the bending angle by an Ellis wormhole goes as $1/u^2$ instead of $1/u$ (where $u$ is the impact parameter) \cite{EllLen,Abe,EllLen2} and that it is possible to obtain a total amplification smaller than one. This implies that for certain geometries the combined flux of the images is smaller than the original flux of the source (defocusing effect), which is impossible to achieve in ordinary gravitational lensing by isolated objects \cite{Abe} (in the context of ordinary matter, demagnification can only occur for lines of sight along underdense regions \cite{Hilbert09}). Actually, defocusing is not an accident of the Ellis wormhole metric. Gravitational lensing by metrics falling as $1/r^n$ has been studied by Kitamura et al. \cite{Kit2013}, who showed that the deflection angle inherits the same exponent as the metric: $\hat\alpha\sim 1/u^n$. They also showed that the defocusing effect is generic for any metrics falling as $1/r^n$, with $n>1$. The physical problem of gravitational lensing by metrics falling as $1/r^n$ has been deeply investigated in a sequence of papers appeared in the last few years \cite{AsadaAll}. Metrics falling as  $1/r^n$ were first studied in Ref. \cite{Kit2013} as a phenomenological extension of the Ellis wormhole case, while the implications for the energy-momentum tensor supporting this kind of metrics were only considered in Ref. \cite{BozPos}. They prove that some violation of the weak energy density must necessarily occur in order to have a metric falling faster than Schwarzschild ($n>1$). As a consequence, the observation of defocusing in some physical lenses would provide a smoking gun for the existence of exotic matter. On the other hand, many authors have proved that it is possible to obtain wormhole solutions without exotic matter in modified theories of gravity \cite{WorMod}.

With all these recent studies motivating further investigation on the observational properties of non-Schwarzschild metrics, with the present paper we aim at extending the existing gravitational lensing theory for $1/r^n$ metrics developed in Refs. \cite{Kit2013,AsadaAll} to the binary case. Indeed, the defocusing property common to such metrics and the non-trivial structure of the Jacobian matrix found in the single-lens case incites to make the leap to a case in which the spherical symmetry is broken and the critical structure of the lens map becomes non-trivial. Do the caustics in an exotic binary lens have any resemblance to the standard Schwarzschild case? Is the number of images the same? Do we find any new mathematical features that cannot be reproduced in the standard case? There are many ways to break the spherical symmetry. We can consider the rotation of the source distribution \cite{Rotation}, the presence of an external shear \cite{ChaRef}, a binary system composed of a normal star orbiting an exotic object or a binary system composed of two exotic condensates. Indeed, any of these routes require its own amount of work, so, as a first case study, we have decided to focus on the binary system composed of two exotic objects. This kind of system has the advantage of being mathematically simpler than the mixed case (normal star - exotic object) since the two deflection terms come with the same power of the impact parameters. The absence of any standard terms in the lens equation also ensures that any new phenomena can be ascribed to the $1/r^n$ gravity without any contamination from other terms. So, it represents an ideal case in which we can explore a fully developed non-trivial caustic structure generated in a genuine $1/r^n$ framework. Besides mathematics, an astrophysical motivation for studying binary wormholes or $1/r^n$ objects comes from studies of dynamical formation of wormholes from an initial scalar field configuration \cite{Maeda09}. It is likely that local inhomogeneities in this field may cause the formation of multiple objects in the same way as a gas cloud collapses by Jeans instability generating stellar clusters. So, a wide-range investigation of exotic objects should take into account the possibility that they form in binaries or clusters.

The standard binary lens caustic structure has been studied by Schneider and Weiss \cite{SchWei} and by Erdl and Schneider \cite{ErdSch}. They showed that, depending on the projected distance between the two components, three different topology regimes exist. Interesting results on the magnification pattern were obtained by Witt and Mao \cite{WitMao} and Asada \cite{Asada02}. Getting more closely to our work, Shin and Evans discussed the critical curves and caustics of isothermal haloes with and without a regular core \cite{ShinEvans}. Notably, the singular isothermal sphere provides a limiting case for the $1/r^n$ objects obtained for $n\rightarrow 0$. Gravitational potentials with $n<1$ characterize dust (non-relativistic matter) distributions whose 3-dimensional density profile falls as $1/r^{n+2}$. Since haloes of galaxies may have different profiles depending on the amount of dark matter they contain, our study has interesting applications to lensing by galaxy systems. We may e.g. consider the gravitational lens formed by two galaxies with spheroidal haloes declining as $1/r^{n+2}$. The structure of critical curves and caustics is obviously a function of the exponent $n$. A complete understanding of these mathematical curves in different regimes may help modeling these systems and fix the slope parameter in real astrophysical cases. This would be useful to extend existing catalogues of unusual gravitational lenses \cite{OrbMar} beyond the basic isothermal profile.

In definitive, we will have two regimes for our investigation of binary $1/r^n$ objects: $n<1$ corresponding to galactic haloes, $n>1$ corresponding to exotic objects violating energy conditions. The familiar Schwarzschild metric comes with $n=1$ and represents the divider between these two very different classes.

The paper is organized in the following sections. Section 2 reviews $1/r^n$ metrics and the single lens model explored in other papers. Section 3 introduces the binary lens equation for $1/r^n$ objects. In Section 4 we explore the critical curves and the caustics of this lens in all ranges of parameters. In Section 5 we derive the formulae for the boundaries between different topology regimes. In Section 6 we derive the analytical limits of the caustics in the extreme cases (wide binary, close binary, very small strength ratio). We draw our conclusions in Section 7.

\section{Gravitational lensing by non-Schwarzschild objects} \label{Sec single}

Let us introduce the metric \cite{Kit2013}

\begin{equation}
ds^2=\left( 1-\frac{\alpha}{r^n} \right) c^2dt^2-\left( 1+\frac{\gamma}{r^n} \right)\left( dr^2-r^2d\theta^2 + \sin^2\theta d\phi^2 \right), \label{metric}
\end{equation}

depending on the coefficients $\alpha$ and $\gamma$. As shown in Ref. \cite{BozPos}, in order to support such a metric, Einstein equations require precise relations between these coefficients and the energy-momentum tensor components. In particular, we recall that the exponent $n$ only depends on the ratio between tangential and radial pressure $n=-2 p_t/p_r$.

The deflection of light in such a metric was first studied by Kitamura et al. \cite{Kit2013}. They find
\begin{equation}
\hat\alpha=\frac{\sqrt{\pi}\Gamma[(1+n)/2]}{\Gamma[n/2]} \frac{\alpha+\gamma}{u^n}, \label{Angle}
\end{equation}
where $\Gamma[z]$ is the Euler gamma function and $u$ is the impact parameter of the light ray.

The strength of the deflection is directly proportional to the sum $\alpha+\gamma$. More in general, following Ref. \cite{KopSch}, the deflection from a generic energy-momentum distribution is
\begin{equation}
\hat{\vec \alpha}(\vec x)=\frac{4G}{c^4}\int d^2x'\frac{\vec x-\vec x'}{|\vec x-\vec x'|^2}  \int dz \left(\rho+p_z\right),
\end{equation}
where the photon unperturbed path is along the $z$ axis and $p_z$ is the pressure component along this direction. Light deflection is sensitive both to the energy density and the pressure, which is non-negligible in our case. So, when introducing the binary lens case, we will not refer to a ``mass ratio'' between the two components, but rather to a ``strength ratio''. This terminology perfectly applies to the ordinary matter $n<1$ case, where the coefficients are determined by the local density normalizations rather than the total masses of the haloes.

The lens equation is then
\begin{equation}
\beta=\theta - \frac{\theta_E^{n+1}}{|\theta|^n}\mathrm{Sign}(\theta),
\end{equation}
where $\beta$ is the source angular position with respect to the center of the lens, $\theta=u/D_L$ is the angular position at which the observer observes the image and the modified angular Einstein radius reads
\begin{equation}
\theta_E=\left[\frac{D_{LS}}{D_L^nD_S}\frac{\sqrt{\pi}\Gamma[(1+n)/2]\left( \alpha+\gamma \right)}{\Gamma[n/2]} \right]^{1/(n+1)}.
\end{equation}

For $n>0$ there are still two real images, whose position can be calculated numerically. In the limit $\beta \gg \theta_E$, the primary image is in $\theta_+\simeq \beta$ and the secondary image is in $\theta_-\simeq - (\theta_E^{n+1}/\beta)^{1/n}$. In the singular isothermal sphere ($n=0$), the secondary image can only exist when the source is within the Einstein radius $\beta<\theta_E$. The emergence of a single image from the crossing of the Einstein ring clearly violates the theorems on the number of images. This is a consequence of the presence of the central singularity.

The radial and tangential magnifications are respectively given by
\begin{eqnarray}
&& \mu_r=\left(\frac{\partial \beta}{\partial \theta} \right)^{-1}= \left( 1+n \frac{\theta_E^{n+1}}{\theta^{n+1}}\right)^{-1} \\
&& \mu_t=\frac{\theta}{\beta}=\left( 1- \frac{\theta_E^{n+1}}{\theta^{n+1} }\right)^{-1},
\end{eqnarray}
with the total magnification given by $\mu=\mu_r\mu_t$.

As noted in Ref. \cite{Kit2013}, for $n>1$ there always exists some value of $\beta$ leading to a total magnification smaller than one (defocusing). This can be simply obtained by an expansion of $\mu$ for $\theta \simeq \beta \gg \theta_E$ and neglecting the contribution by the secondary image:
\begin{equation}
\mu \rightarrow 1 +(1-n) \left(\frac{\theta_E}{\beta}\right)^{n+1}+(1-n+n^2) \left(\frac{\theta_E}{\beta}\right)^{2n+2} + o\left(\frac{\theta_E}{\beta}\right)^{2n+2}.
\end{equation}
The first deviation from unity is negative for any $n>1$. For the Schwarzschild case, the leading deviation is the second order, scaling as $\beta^{-4}$ as well known. In practice, images created by exotic lenses are as extended as Schwarzschild ones tangentially, but are more compressed on the radial direction. While the two deformations compensate at first order in Schwarzschild, this no longer happens for exotic lenses. For what concerns ordinary matter distributions ($n<1$), the situation is reversed, with a radial compression that is smaller than in the Schwarzschild case, allowing a higher total magnification emerging at the $1/\beta^2$ level already.

When the source is perfectly aligned behind the lens ($\beta=0$), the image coincides with the only tangential critical curve, the Einstein ring with radius $\theta_E$. So, the caustic structure of the single $1/r^n$ lens is analogous to its standard cousin: one tangential critical curve corresponding to the degenerate caustic point in the origin. There are no radial critical curves in the $1/r^n$ lenses studied in this paper. In the singular isothermal sphere ($n=0$), actually the circle with radius $\theta_E$ divides the source plane in a region where we have a single image and an internal region in which we have two images. The name pseudo-caustic has been proposed \cite{EvaWil,ShinEvans} for the curve with radius $\theta_E$. This pseudo-caustic becomes a real radial caustic whenever we allow for a central regular core, where a third highly-demagnified image can live. In principle, radial critical curves can also be obtained in the regime $ n<0 $, which corresponds to density profiles decaying slower than the singular isothermal sphere. This case differs substantially from the regime investigated in this paper, as the deflection angle grows with the impact parameter instead of decreasing.

It is interesting to note that the angular Einstein radius in the standard Schwarzschild lens scales as the square root of the mass. So, if one changes the mass by a factor $q$, the Einstein radius scales as $\sqrt{q}$. In the general $1/r^n$ case, if one changes the energy-momentum distribution by scaling it by a factor $q$, the deflection angle at a fixed impact parameter scales as $q$, but the Einstein radius scales as $q^{1/(n+1)}$, which contains the Schwarzschild limit for $n=1$. This difference has another interesting consequence: while the area of the critical curve is directly proportional to the mass in the Schwarzschild case, this is not true for non-Schwarzschild objects. The area enclosed in the critical curve grows more slowly with the lens strength.

\section{Binary lenses}

Now let us consider a lensing system composed of two objects whose gravitational potential asymptotically falls as $1/r^n$, whether ordinary ($n<1$) or exotic ($n>1$). If the two objects are gravitationally bound, the distribution of the matter-source supporting either object would be warped by the presence of the partner. The resulting distribution could only be obtained by a consistent resolution of the Einstein equations with the hydrodynamical equations of the source. We have no intention to tackle such a complicated problem, since our focus is on all possible new effects on the gravitational lensing sector. So, the only case in which we can consider the total light deflection as the sum of the contributions given by the two isolated objects arises when the two objects stand far apart along the line of sight but perspectivally close enough so as to affect the light trajectory by their peculiar gravitational potential. On the other hand, if we do not want to tangle with multiple plane deflection issues \cite{multiple,Werner08,ColBac}, we must still require that the distance between the pair of lenses is much smaller than the source-lens and lens-observer distances. The mixed case, with a star orbiting the exotic object is probably more physically appealing, but more complicated mathematically. This paper intends to climb a clear and simple first step in this topic.

We then consider two non-Schwarzschild objects A and B whose reduced deflection angles are given by
\begin{equation}
\alpha_{A,B}(\vec\theta) = \epsilon_{A,B} \theta_{E}^{n+1} \frac{\left(\vec\theta-\vec\theta_{A,B}\right)}{|\vec\theta-\vec\theta_{A,B}|^{n+1}}.
\end{equation}

The coordinates in the sky of the two objects are $\vec\theta_{A,B}$, the deflection strengths depend on the gravitational potentials of the two objects through Eq. (\ref{Angle}). Setting $\epsilon_A+\epsilon_B=1$, the strength ratio is $q\equiv \epsilon_A/\epsilon_B$, while the order of magnitude of the deflection is set by the angular scale $\theta_{E}$. This corresponds to the angular radius of the Einstein ring of an isolated object with $\epsilon_i=1$. As emphasized before, for each object, this radius scales with $\epsilon_i^{n+1}$.

At this point, we can write down the lens equation taking these two contributions into account
\begin{equation}
\vec\beta=\vec\theta-\epsilon_{A} \theta_{E}^{n+1} \frac{\left(\vec\theta-\vec\theta_{A}\right)}{|\vec\theta-\vec\theta_{A}|^{n+1}}-\epsilon_{B} \theta_{E}^{n+1} \frac{\left(\vec\theta-\vec\theta_{B}\right)}{|\vec\theta-\vec\theta_{B}|^{n+1}}. \label{LensEq}
\end{equation}

For the ordinary binary lens equation ($n=1$) it is very useful to introduce complex coordinates, as first envisaged by Witt \cite{Witt}. We do the same in the generic power $n$ case as follows
\begin{equation}
\zeta =\frac{\beta_{1}+ i \beta_{2}}{\theta_{E}}; \; \; z =\frac{\theta_{1}+ i \theta_{2}}{\theta_{E}}.
\end{equation}

For simplicity, we take the origin of coordinates at the mid-point between the two lenses $  z_{A}=-z_{B}$ with the real axis along the line joining the lenses. We call the normalized angular separation between the lenses $ s $, so that we can set $ z_{A }= -s/2 $ and $ z_{B} = s/2 $.
Then the lens equation (\ref{LensEq}) can be written as
\begin{equation}
\zeta=z-\frac{\epsilon_{A}}{(z+s/2)^{\frac{n-1}{2}}(\bar{z}+s/2)^{\frac{n+1}{2}}}-\frac{\epsilon_{B}}{(z-s/2)^{\frac{n-1}{2}}(\bar{z}-s/2)^{\frac{n+1}{2}}}.
\end{equation}

The fortunate simplification that removes $z$ from the deflection terms only occurs for $n=1$. Its presence in the general case severely complicates the Jacobian determinant of the lens map, which in complex notations is given by
\begin{equation}
J(z, \bar{z})=\left|\dfrac{\partial \zeta}{\partial z}\right|^{2}-\left|\dfrac{\partial \zeta}{\partial \bar z}\right|^{2}.
\end{equation}

In fact, the first term, which was just unity in the Schwarzschild case, now becomes as involved as the second. In detail, the Jacobian is
\begin{eqnarray}
J & =\left\lbrace 1 + \dfrac{n-1}{2}\left[ \dfrac{\epsilon_{A}}{(z+s/2)^{\frac{n+1}{2}}(\bar z +s/2) ^{\frac{n+1}{2}}}+\dfrac{\epsilon_{B}}{(z-s/2)^{\frac{n+1}{2}}(\bar z -s/2) ^{\frac{n+1}{2}}} \right] \right\rbrace ^{2}\nonumber \\
& \quad -\dfrac{(n+1)^{2}}{4}\left\vert  \dfrac{\epsilon_{A}}{(z+s/2)^{\frac{n+3}{2}}(\bar z +s/2) ^{\frac{n-1}{2}}}+\dfrac{\epsilon_{B}}{(z-s/2)^{\frac{n+3}{2}}(\bar z -s/2) ^{\frac{n-1}{2}}} \right\vert^{2}. \label{Jac}
\end{eqnarray}

The critical curves of a lens model are defined by the condition $J(z)=0$ \cite{SEF}. Creation/destruction of the images can only occur on such points. The corresponding points on the source plane, found by applying the lens map, form the caustic curves. Caustic curves bound regions in which a source gives rise to an additional pair of images. A source on a caustic point produces (at least) two degenerate images on the corresponding critical point. For all these properties, critical curves and caustics are generally sufficient to understand the qualitative behavior of a lens system. They are widely used for the interpretation of observations in all applications of gravitational lensing.

Our goal, achieved in the following sections, is a full numerical and analytical exploration of critical curves and caustics of the generalized binary lens with deflections falling as $1/u^n$.

\section{Numerical exploration of Critical curves and Caustics}

We start from the symmetric equal-strength case and then move to the asymmetric strength case. Following the notation introduced in the previous section, we assume that the centers of the two lenses lie on the real axis (using complex notations) and we choose the origin half-way between the two lenses. The unit of measure is $\theta_E$, i.e. the angular radius of the Einstein ring that we would obtain for an isolated object with strength $\epsilon=1$. The strengths ratio of the two object is $q= \epsilon_A/\epsilon_B$ and their separation in units of $\theta_E$ is $s$. In order to show the evolution from $n=1$ to higher values of $n$, we decided to plot the critical curves and caustics for $n=0,0.5,1,2,3$. $n=1$ corresponds to two ordinary Schwarzschild lenses \cite{SchWei,ErdSch}, while $n=2$ corresponds to two Ellis wormholes. A generic value of $n$ can be obtained in other cases, as explained e.g. in Ref. \cite{BozPos}. On the other side, galactic haloes can be obtained with this model in the range $0<n<1$, with the singular isothermal sphere being the limit $n=0$.

The technique used to obtain the critical curve is accurate contouring of the Jacobian determinant using \texttt{Mathematica}. The contours are then mapped through the lens equation to obtain the corresponding caustics in the source plane.

\subsection{Equal-strength binary $q=1$}

In the standard Schwarzschild binary lens, it is well-known that three topologies exist, depending on the separation of the two masses \cite{SchWei,ErdSch}. The close topology is obtained for $s<1/\sqrt{2}$. It consists of a primary 4-cusped caustic at the barycenter of the system generated by the common critical curve enveloping the whole lens and a pair of small triangular caustics generated by two symmetrical critical ovals close to the lenses containing local maxima of the Jacobian. The intermediate topology exists in the range $1/\sqrt{2}<s<2$ and is characterized by a single 6-cusped caustic obtained when the two secondary caustics of the close topology merge with the primary one removing two pairs of cusps by a beak-to-beak singularity. Finally, the wide topology for $s>2$ emerges when the intermediate caustic splits into two 4-cusped astroids along the lens axis; the two astroids become smaller and smaller while the two critical curves tend to the Einstein rings of the two isolated objects.

For the singular isothermal sphere, the same topologies are found \cite{ShinEvans}, but with important additional features in the close regime. In fact, the small ovals shrink to zero size at finite separation $s$ and then grow to finite size again (this singularity is called elliptic umbilic). Finally, these ovals touch the two singular lenses, giving rise to degenerate caustics.

We always find the same three topology regimes in the generic $n$ case, nicely joining and generalizing the two known limits, but with notable differences in the sizes and the shapes of critical curves and caustics, which can be heavily deformed with respect to the known limits.

\begin{figure}[t]
\centering
\includegraphics[width=6cm]{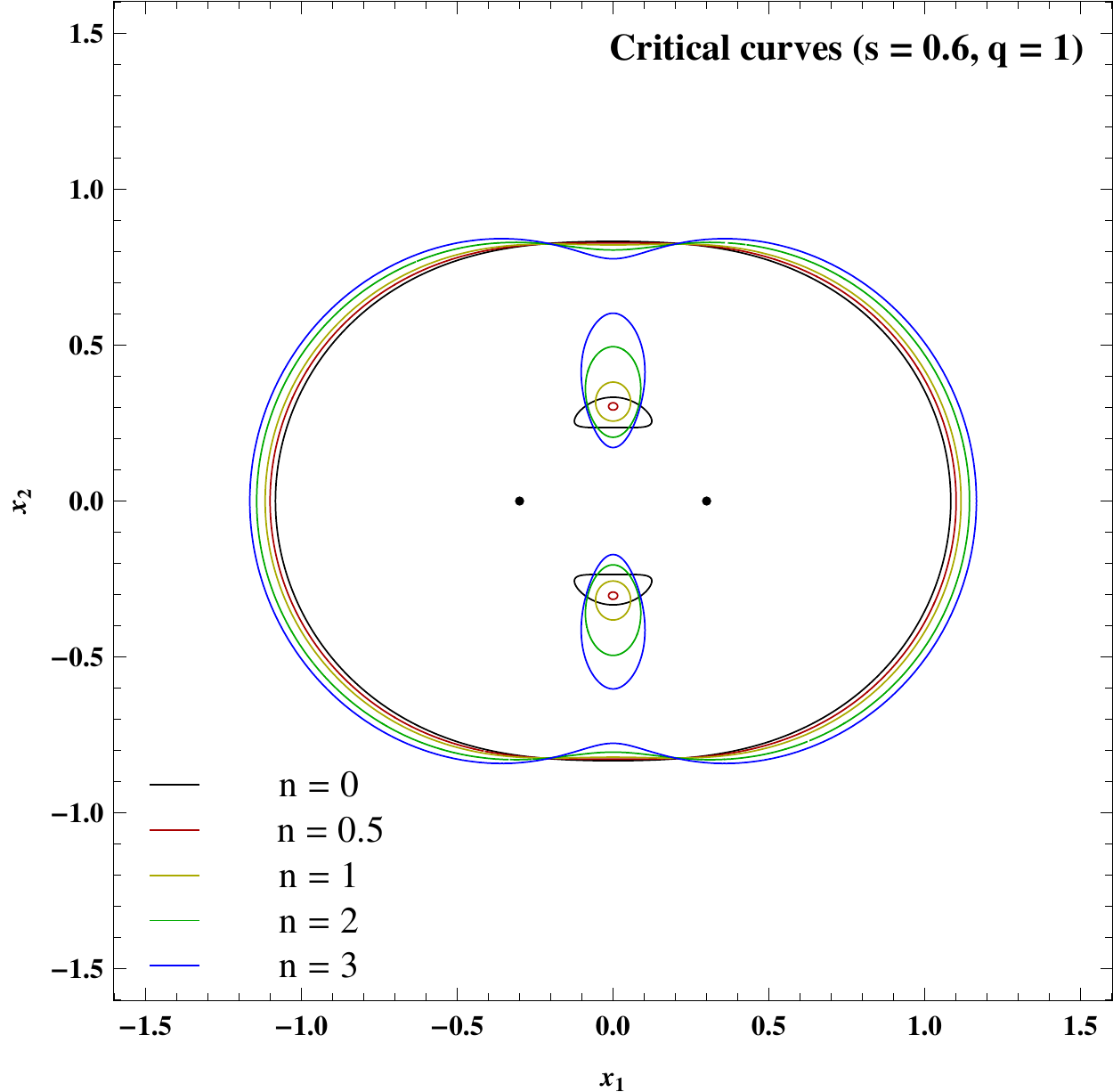}\hspace{4 mm}\includegraphics[width=6cm]{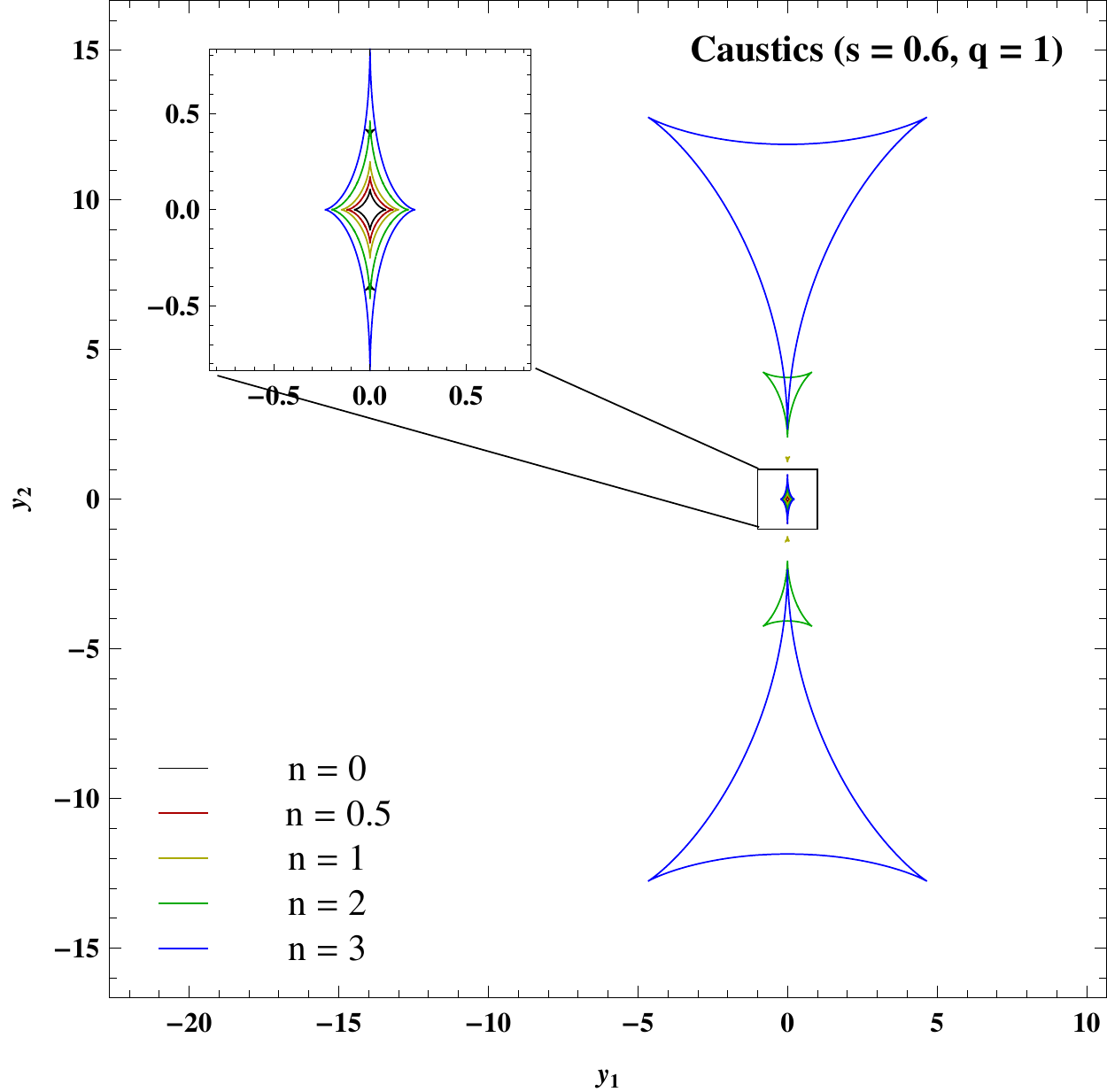}
\caption{Equal-strength binary. Close separation} \label{1c}
\end{figure}

In Fig. \ref{1c} we show the critical curves and caustics for $s=0.6$, which is in the close regime for all values of $n$ chosen. In the standard $n=1$ case, the close topology features three critical curves, one resembling the Einstein ring of the lens obtained by summing the two masses and two smaller critical curves inside the main one. Correspondingly, there is a central astroidal caustic generated by the main critical curve and two triangular caustics corresponding to the small critical curves.

For the singular isothermal sphere ($n=0$), the two secondary critical curves become tangentially elongated and the two caustics become smaller and closer to the primary one. As anticipated, for separations closer than $s=0.25$, the two critical curves reach the two lenses, and the secondary caustics merge with the circular pseudo-caustics \cite{ShinEvans}. The singular isothermal sphere represents a limiting case for our analysis. Indeed, this degenerate behavior is only found for $n$ exactly equal to 0, but for any $n>0$ the critical curves never touch the lenses.

\begin{figure}[t]
\centering
\includegraphics[width=12cm]{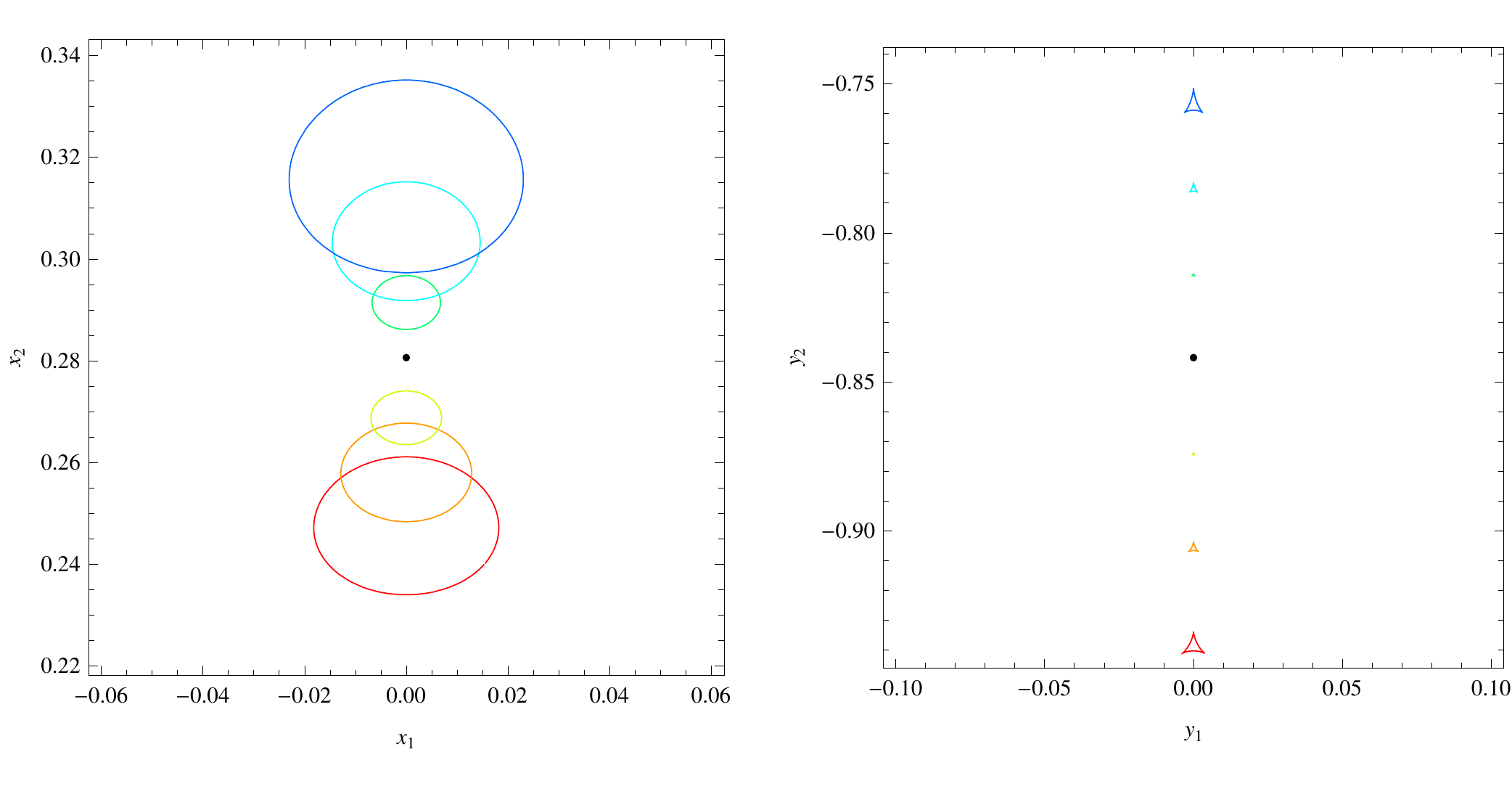}
\caption{The elliptic umbilic catastrophe shown for $n=0.5$ and separations growing from $s=0.5$ to $s=0.62$ in steps of 0.02 from red to blue. The elliptic umbilic occurs at $s=0.56$.} \label{Fig umb}
\end{figure}

On the other hand, we find that the elliptic umbilic catastrophe distinguishing the singular isothermal sphere from the Schwarzschild case is present in the whole range $0\leq n < 1$. Since this feature is particularly difficult to catch in general plots, in Fig. \ref{Fig umb} we propose a sequence of critical curves and caustics at fixed $n=0.5$ and growing separations. From $s=0.5$ to $s=0.56$ the oval (and so the corresponding caustic) shrinks to zero size and then from $s=0.56$ to $s=0.62$ it opens up again. We will come back to the description of this elliptic umbilic in the next section.

\begin{figure}[t]
\centering
\includegraphics[width=6cm]{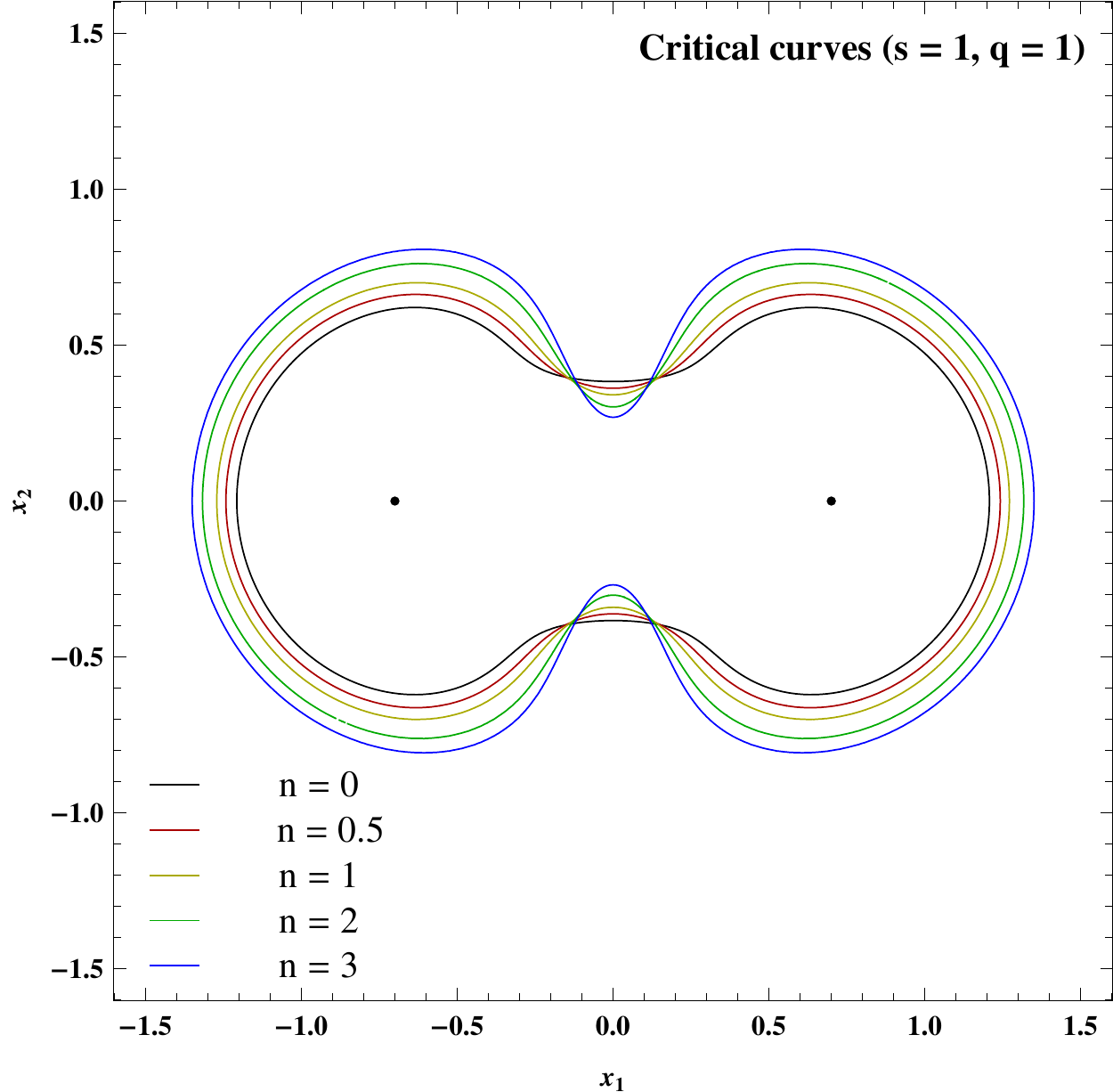}\hspace{4 mm}\includegraphics[width=6cm]{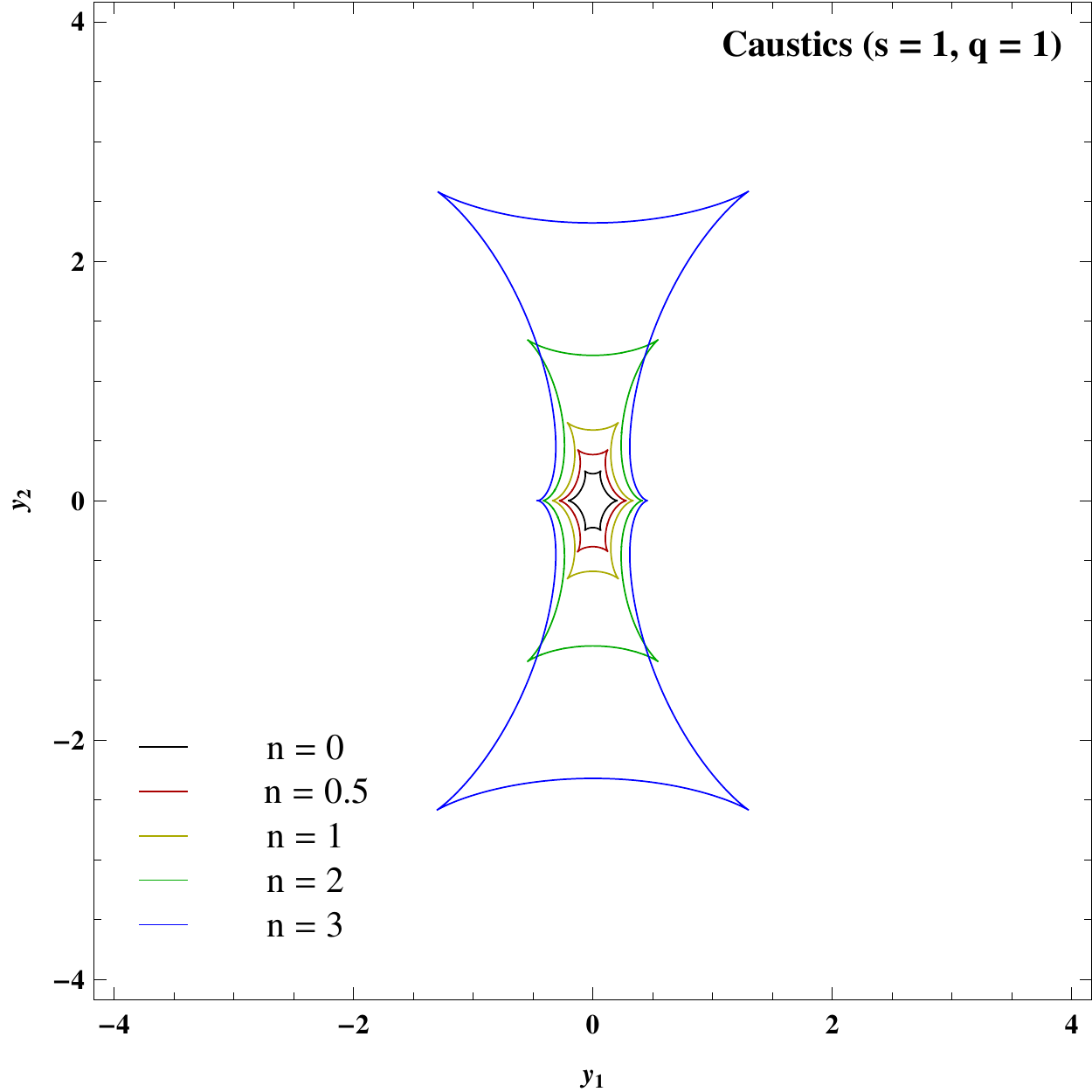}
\caption{Equal-strength binary. Intermediate separation}  \label{1i}
\end{figure}

Moving from 1 to higher $n$ values, we still see the same three critical curves, but the inner ones are severely elongated in the radial direction opposite to the barycenter of the system. In the caustics, this difference causes the formerly small triangular caustics to become bigger and bigger, quickly becoming huge in size (note the scale of the right panel of Fig. \ref{1i}). The consequence of this difference is that it is much easier for exotic lenses to have a source inside one of these giant caustics and then have two additional images close to the corresponding critical curve.

\begin{figure}[t]
\centering
\includegraphics[width=6cm]{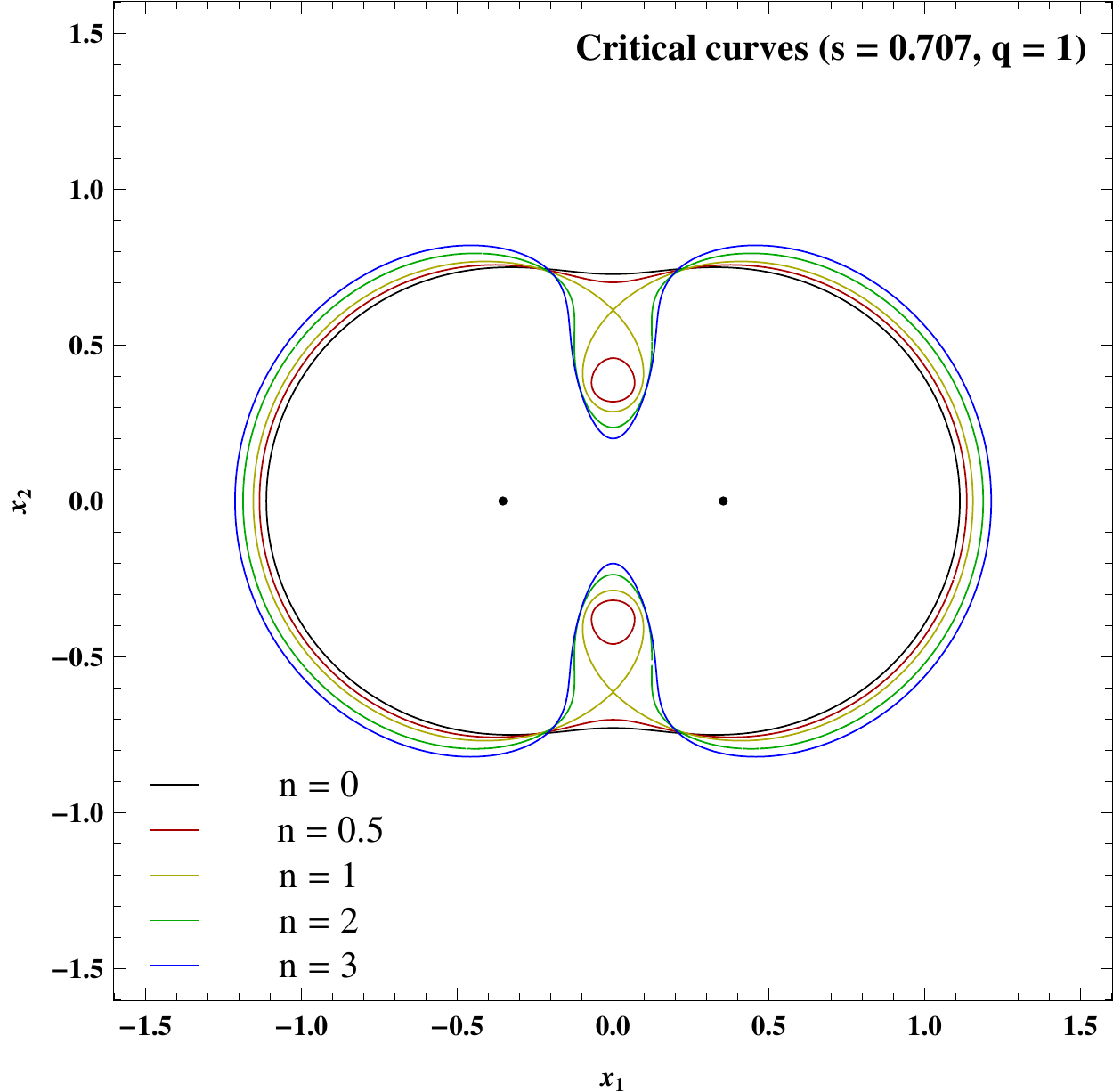}\hspace{4 mm}\includegraphics[width=6cm]{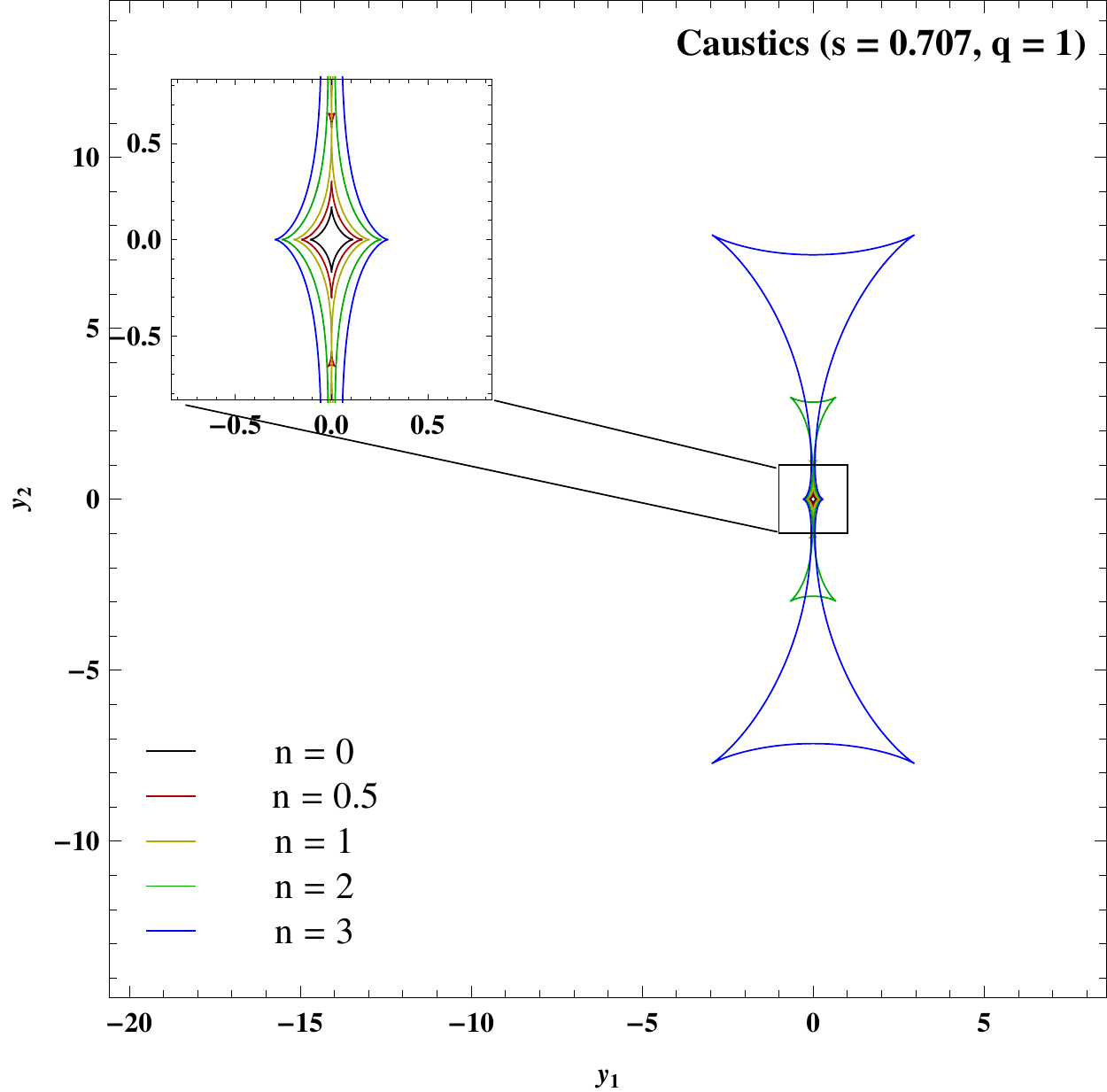}
\caption{Equal-strength binary. Close-intermediate transition} \label{1ci}
\end{figure}

We can imagine that the strong elongation of the inner critical curves will cause the transition from the close to the intermediate regime to occur earlier than in the standard $n=1$ case. Indeed, this is what happens. In fact, in Fig. \ref{1ci} we have chosen $s=1/\sqrt{2}$ to pick up the transition in the $n=1$ case. At this separation, both the $n=2$ and $n=3$ critical curves are already in the intermediate regime, with one single dumbbell-shaped curve and a corresponding 6-cusped caustic. Still the size of these intermediate caustics is much larger than in the standard case. Conversely, the $n<1$ caustics are still in the close regime.

The extension of the intermediate caustics orthogonal to the lens-axis decreases gradually as we go deeper into the intermediate regime. In Fig. \ref{1i}, we still have the $n=2$ and $n=3$ caustics larger than the $n=1$ one, but they are all comparable now. The $n=0.5$ and $n=0$ caustics are still smaller than the $n=1$ one.

\begin{figure}[t]
\centering
\includegraphics[width=6cm]{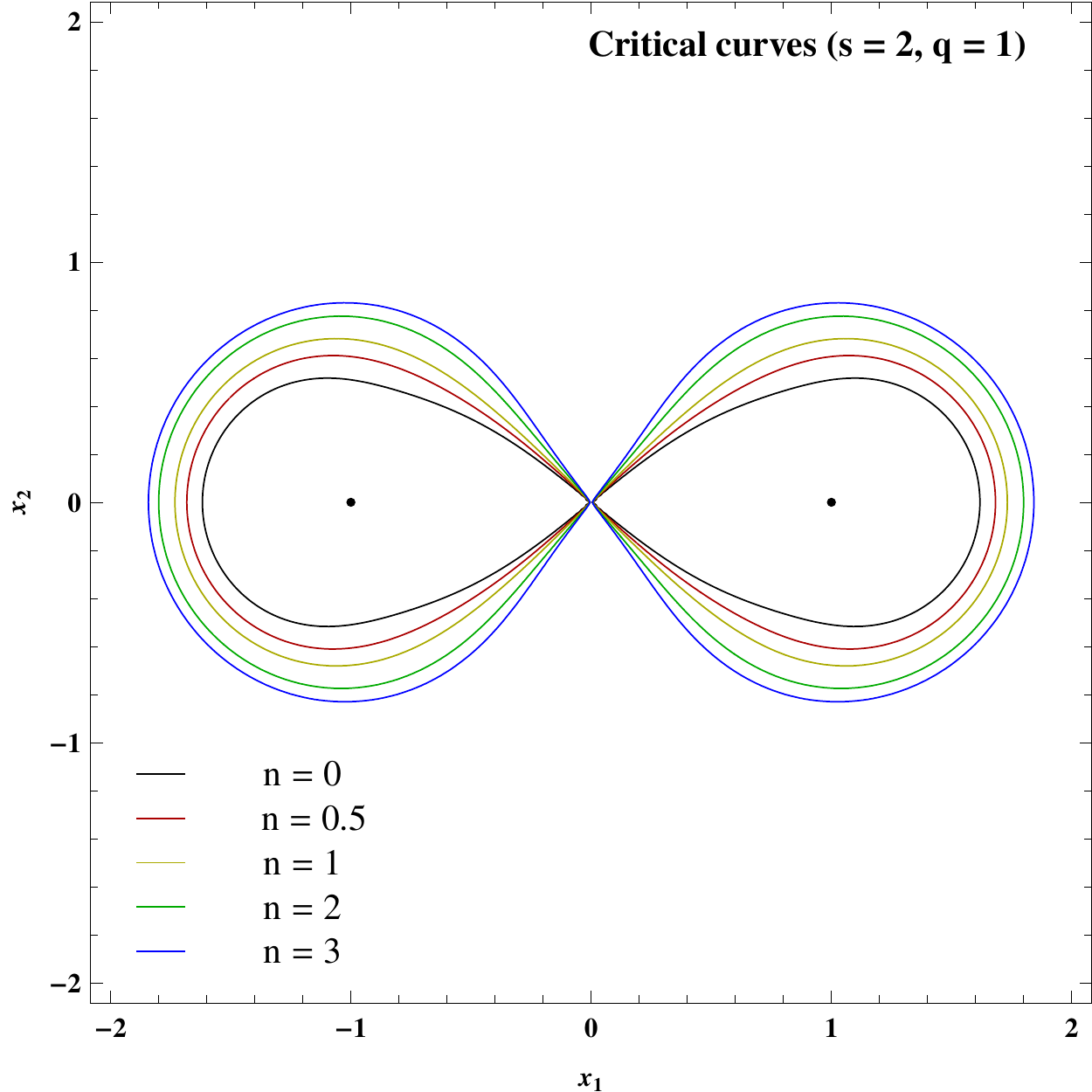}\hspace{4 mm}\includegraphics[width=6cm]{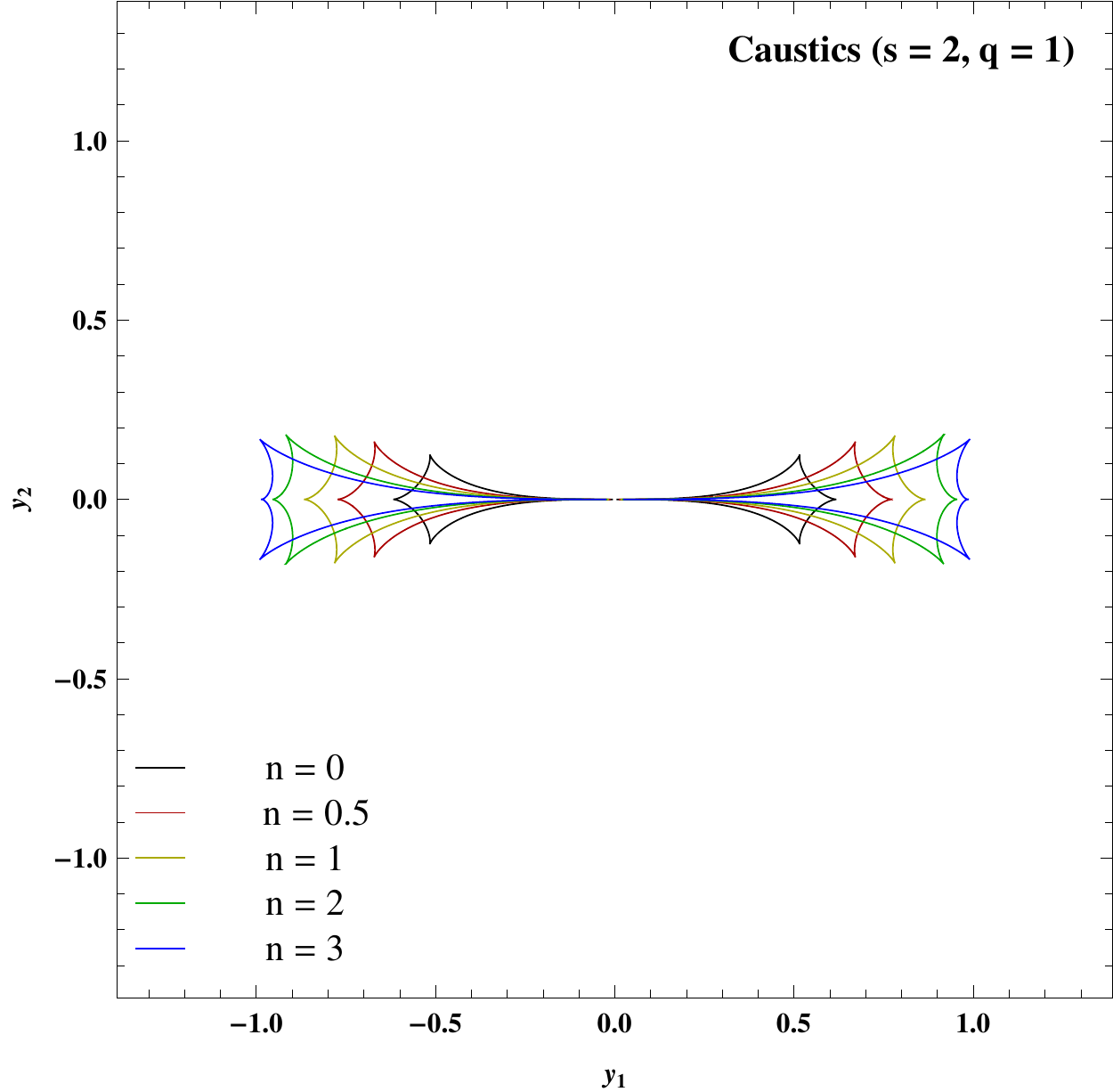}
\caption{Equal-strength binary. Intermediate-Wide transition} \label{1iw}
\end{figure}

The transition between the intermediate and wide regime occurs at $s=2$. This is true not only for $n=1$, but also for all values of $n$, as is evident in Fig. \ref{1iw}. At this separation, the critical curve has a figure-eight shape with the beak-to-beak singularity.

\begin{figure}[t]
\centering
\includegraphics[width=6cm]{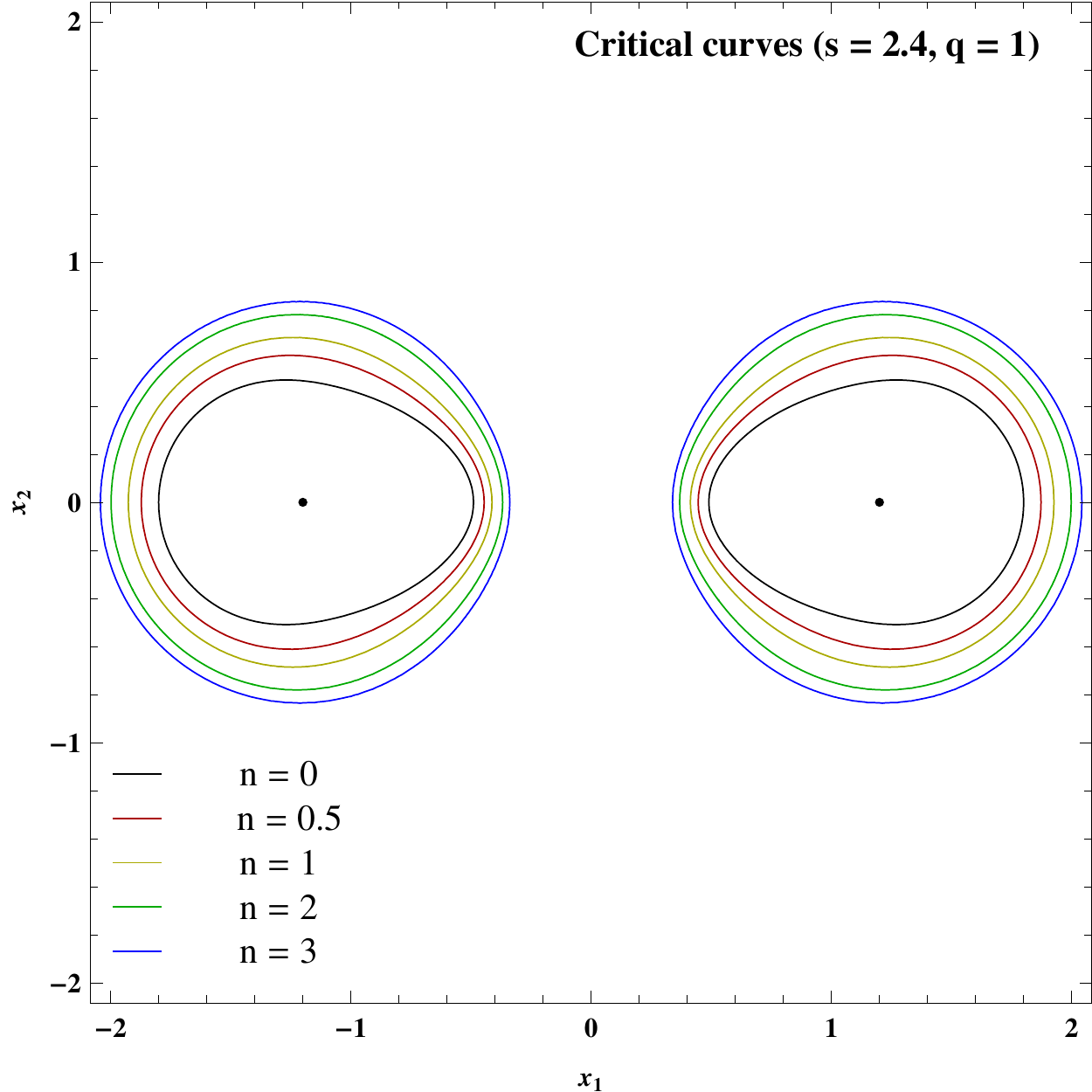}\hspace{4 mm}\includegraphics[width=6cm]{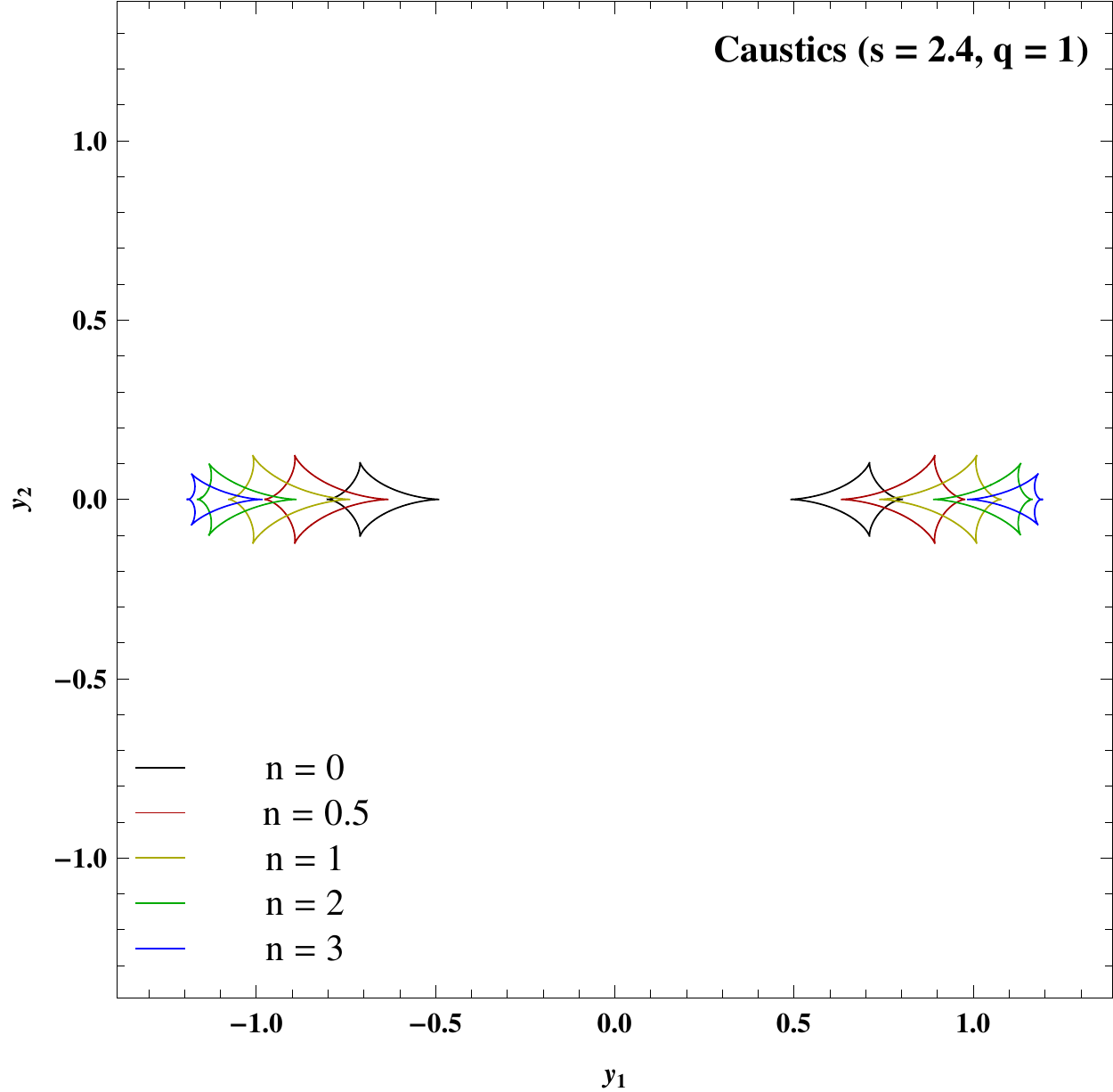}
\caption{Equal-strength binary. Wide separation} \label{1w}
\end{figure}

Finally, Fig. \ref{1w} illustrates the wide regime with $s=2.4$. The critical curves tend to become circular as the lenses move far apart. However, we note that the curves remain more distorted at lower $n$, since the tidal field of the perturbing body decays more slowly. Correspondingly, the largest caustic is now the $n=0$, with the others becoming smaller and smaller at larger $n$. Another thing we should note is that the radius of the critical curves is not the same for all $n$. This is in agreement with the discussion at the end of Section \ref{Sec single}. Since either lens has strength $\epsilon_i=1/2$, the angular Einstein ring of either lens tends to $1/2^{1/(n+1)}$, which grows with $n$.

\subsection{Unequal-strength binary $q=0.1$}

Now we move to the unequal strength case, starting with a moderate strength ratio $q=0.1$.

\begin{figure}[t]
\centering
\includegraphics[width=6cm]{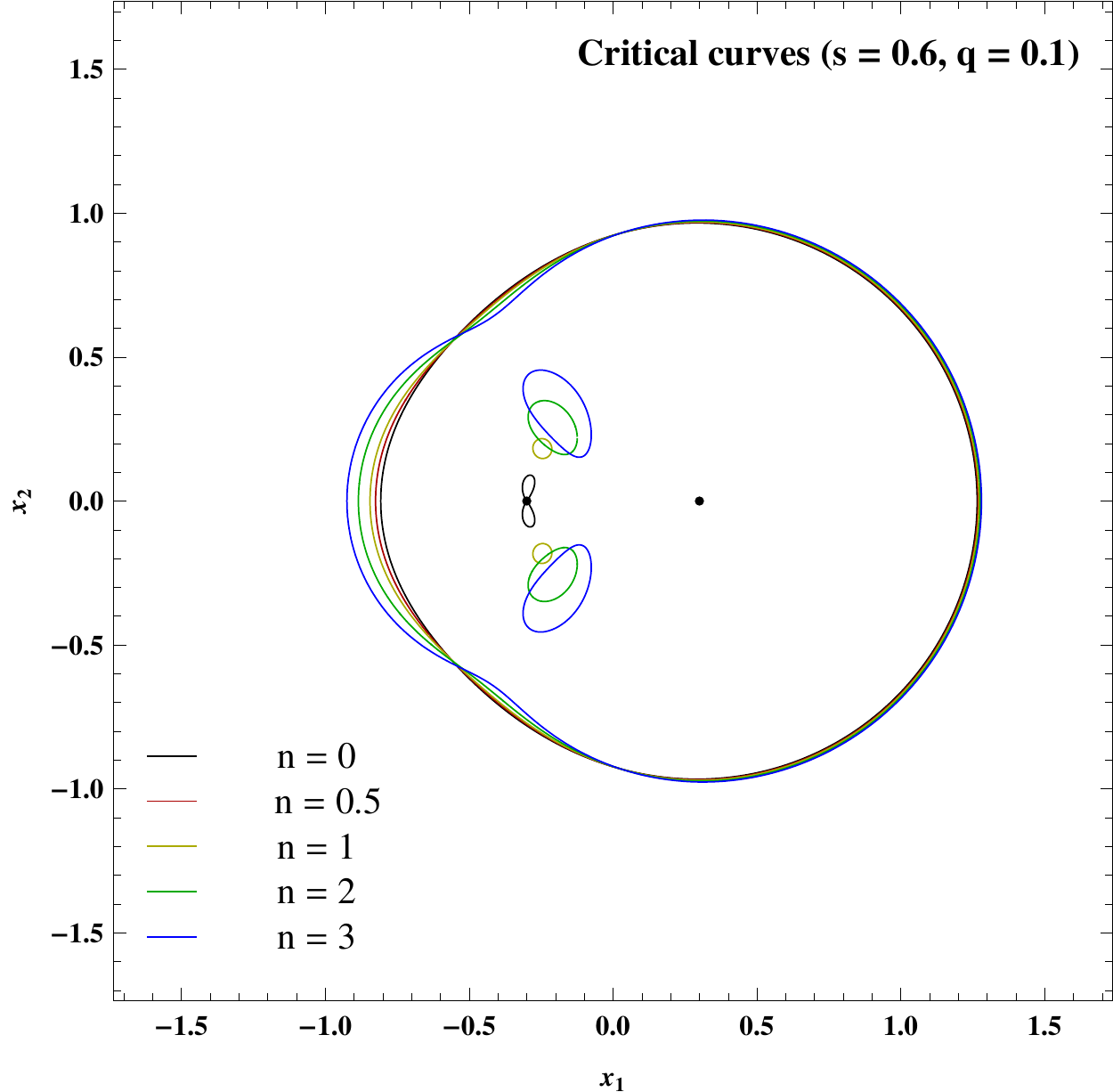}\hspace{4 mm}\includegraphics[width=6cm]{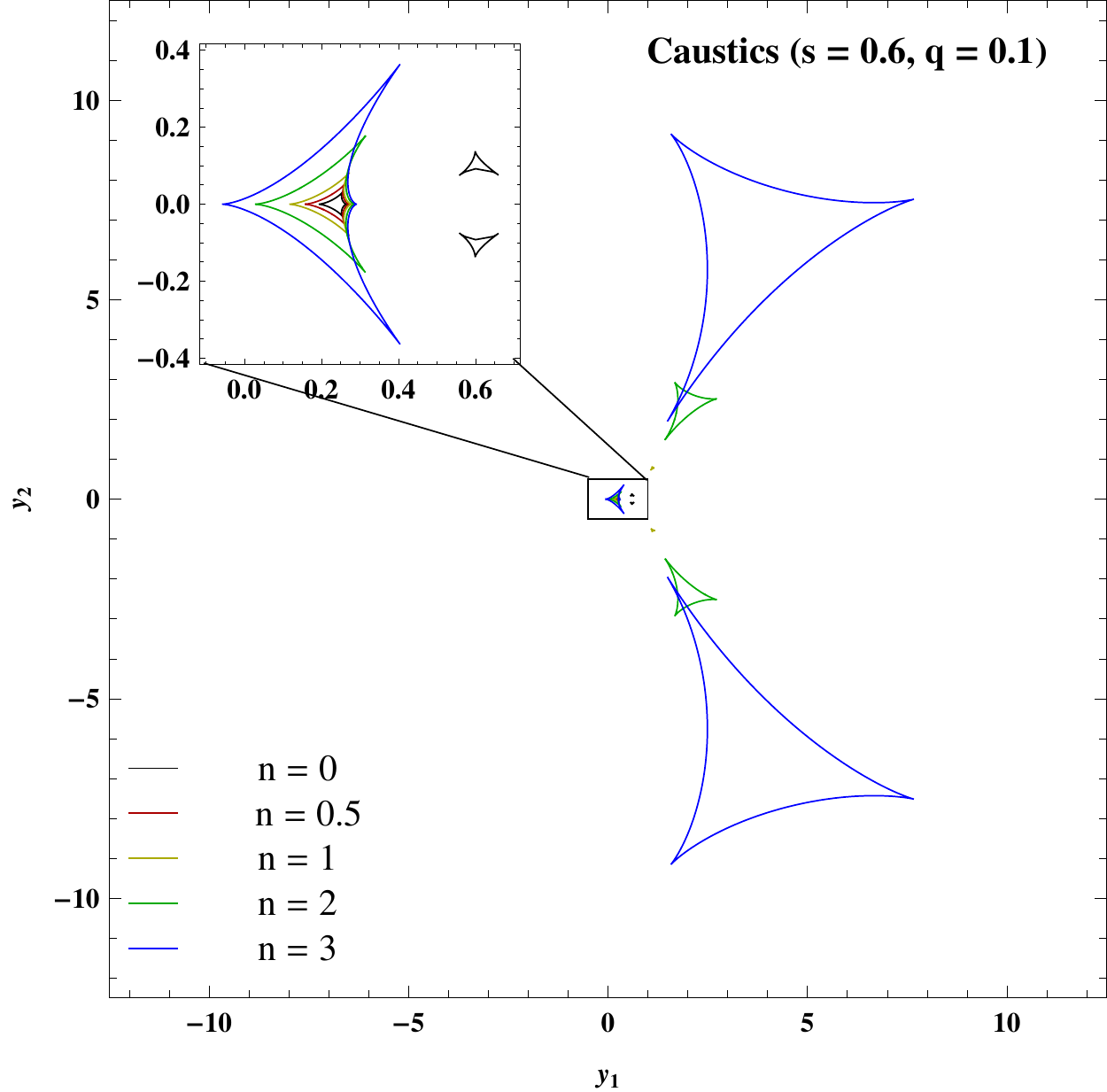}
\caption{Unequal-strength binary. Close separation} \label{2c}
\end{figure}

In the close regime, shown in Fig. \ref{2c}, we see the same characteristics already discussed in the equal-strength case: the main critical curve giving rise to the central caustic and the two smaller critical curves inside the main one, giving rise to the triangular caustics. The asymmetry of the lenses make the secondary critical curves move toward the weaker lens. For ordinary matter distributions ($n<1$), the critical curves rush toward the weaker lens; the secondary caustics become smaller and closer to the primary caustic. Going from the $n=1$-lensing to higher $n$ we see that the inner critical curves become larger, drift toward the center of the system and become radially elongated. The corresponding caustics become larger and larger similarly to the equal-strength case. We also note that the main critical curve is perturbed on the side of the weaker lens, which is again a consequence of the scaling of the individual Einstein rings as $\epsilon_i^{1/(n+1)}$.

\begin{figure}[t]
\centering
\includegraphics[width=6cm]{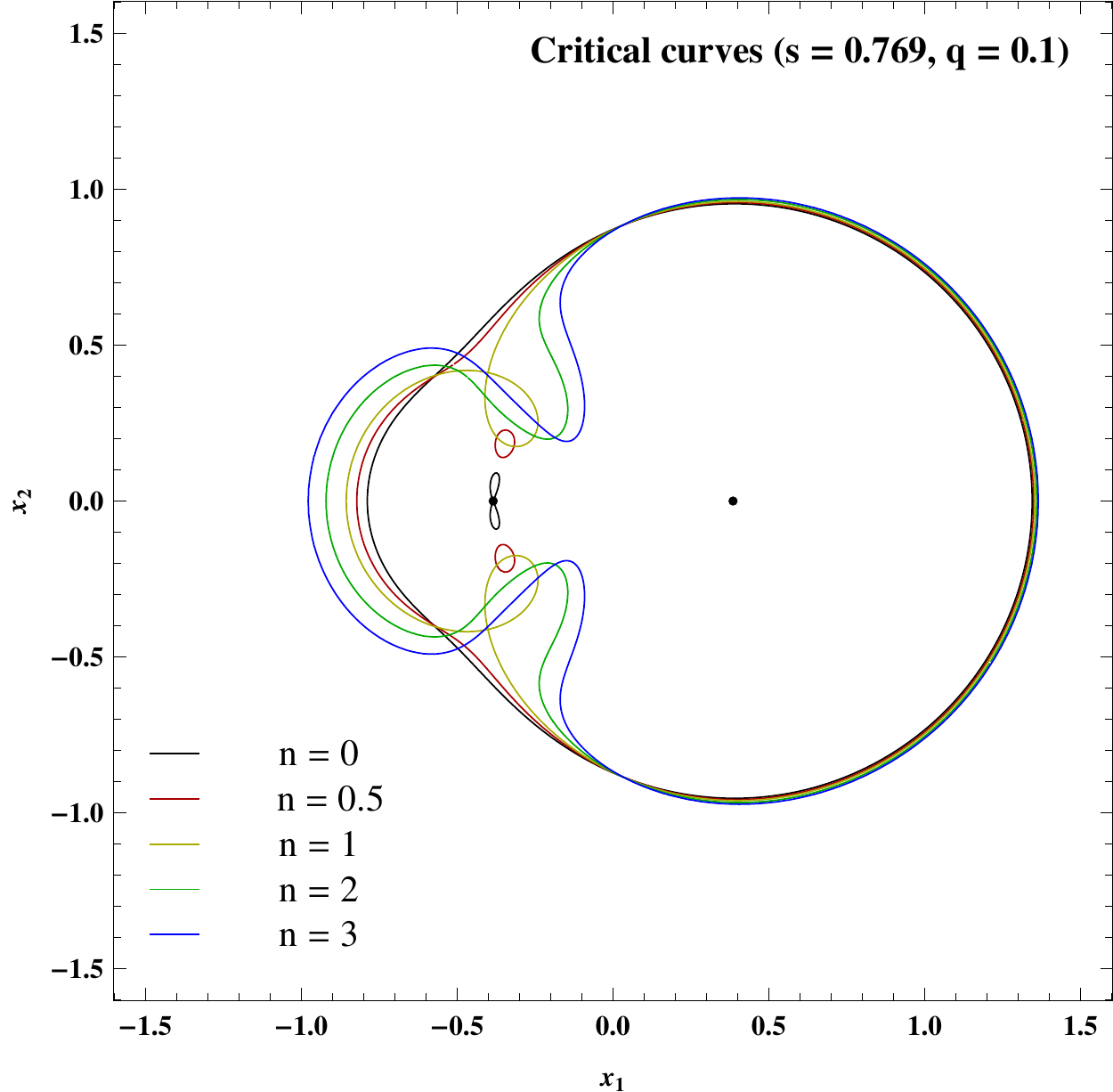}\hspace{4 mm}\includegraphics[width=6cm]{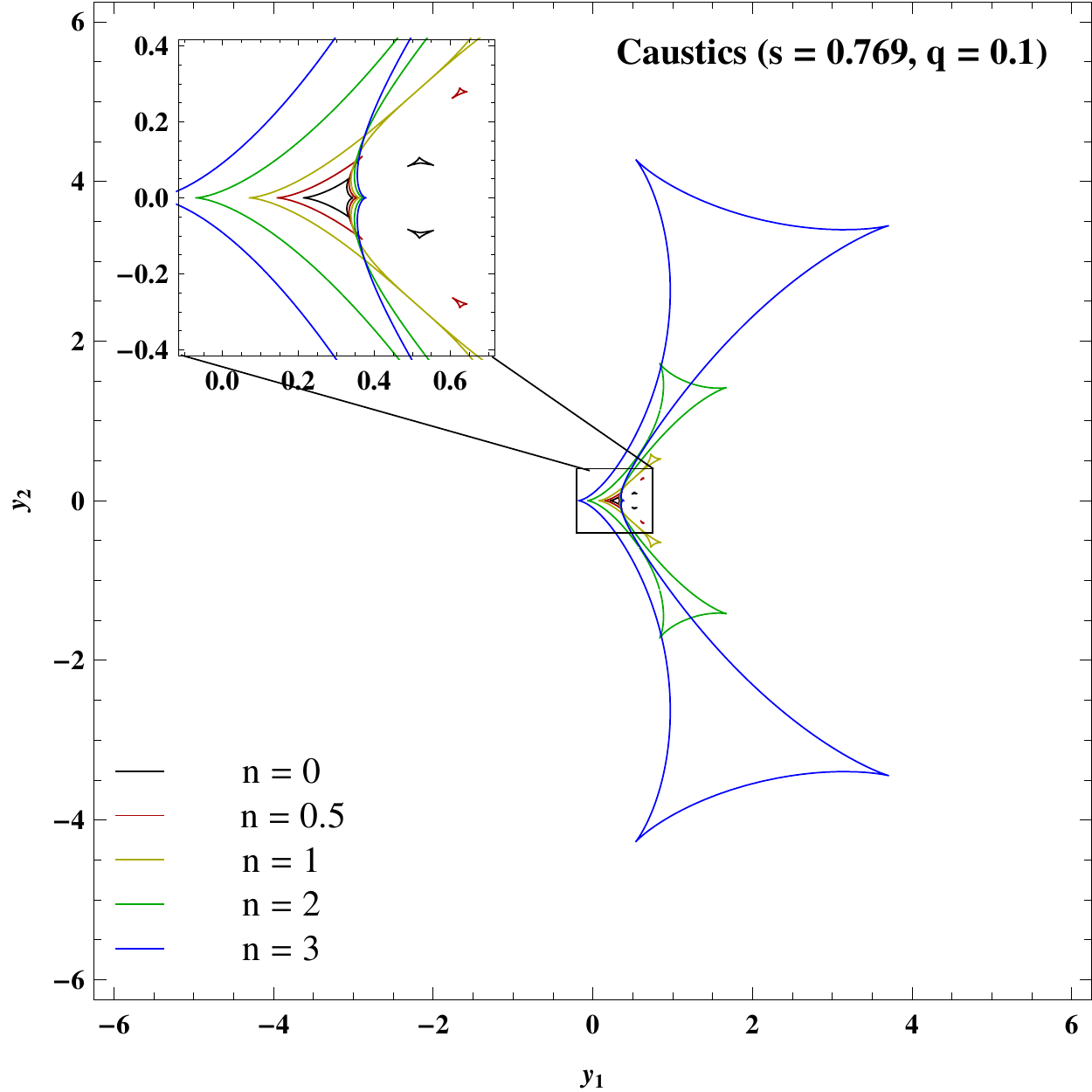}
\caption{Unequal-strength binary. Close-Intermediate transition} \label{2ci}
\end{figure}

In Fig. \ref{2ci} we have $s=0.769$, corresponding to the transition from the close to the intermediate regime in the $n=1$ case. As already found in the equal-strength case, the transition occurs earlier (smaller $s$) for $n>1$ and later (larger $s$) for $n<1$. The $n>1$ caustics are significantly larger than the $n=1$ one.

\begin{figure}[t]
\centering
\includegraphics[width=6cm]{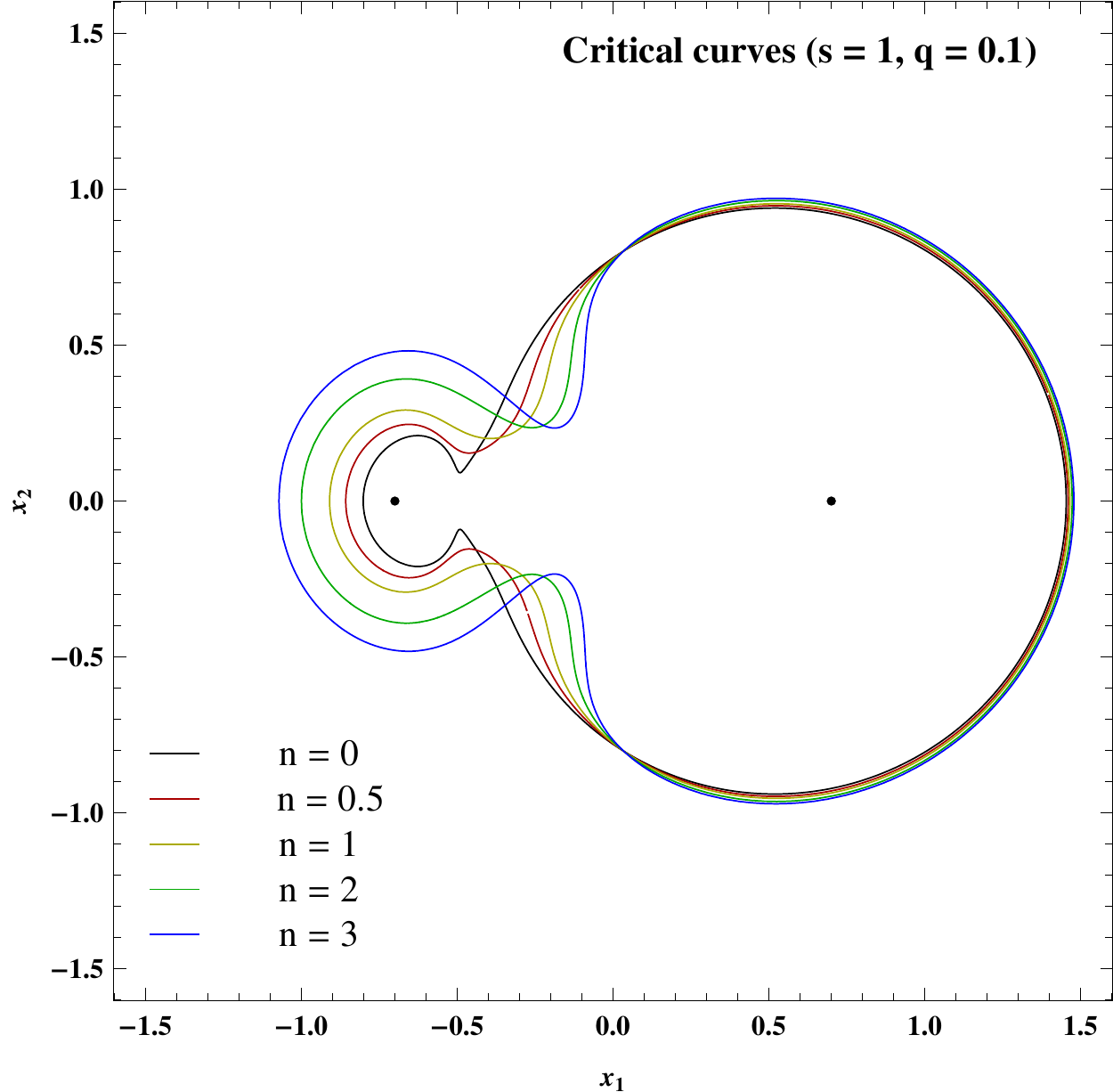}\hspace{4 mm}\includegraphics[width=6cm]{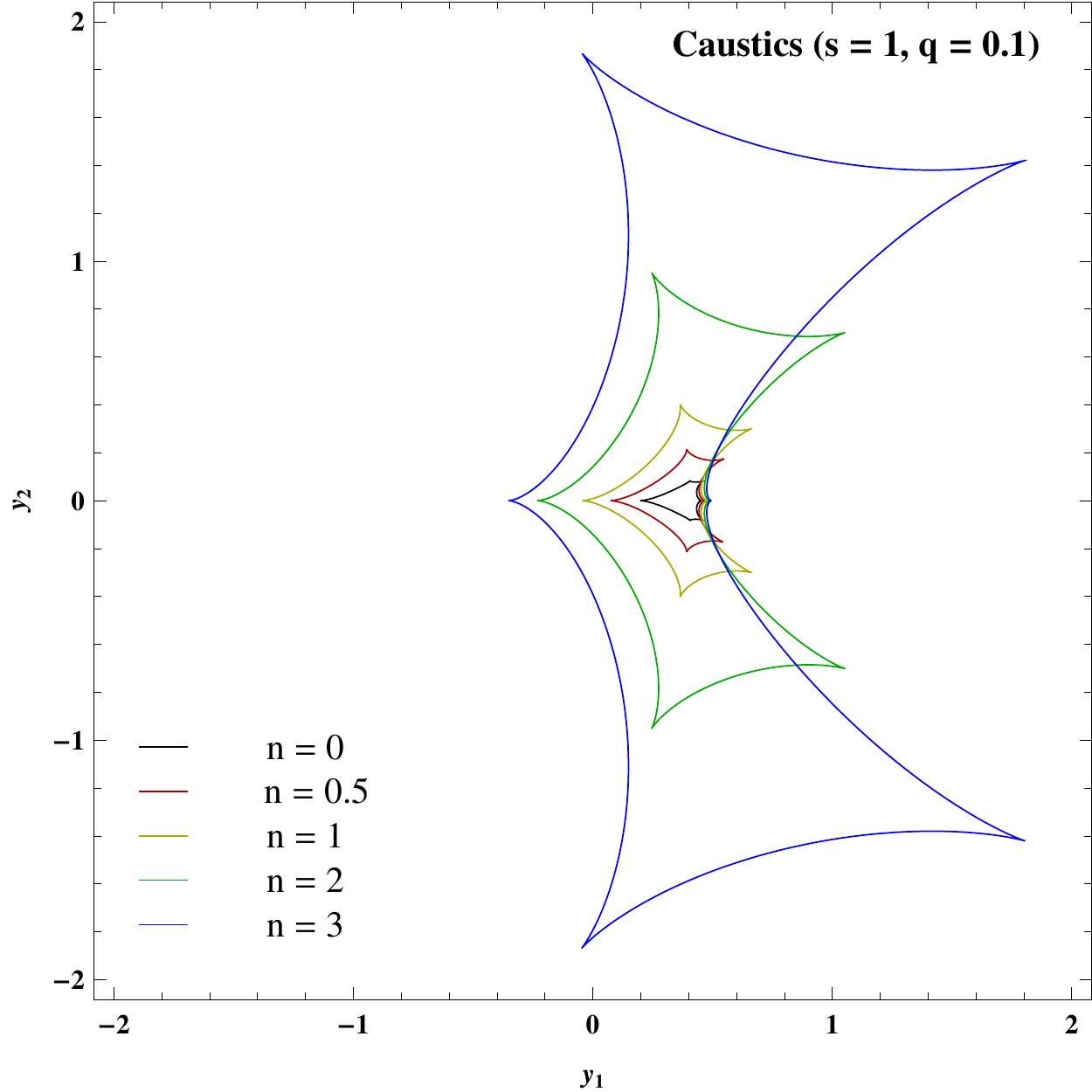}
\caption{Unequal-strength binary. Intermediate separation} \label{2i}
\end{figure}

Then we have Fig. \ref{2i}, with the intermediate regime $s=1$. The caustics become comparable in size.

\begin{figure}[t]
\centering
\includegraphics[width=6cm]{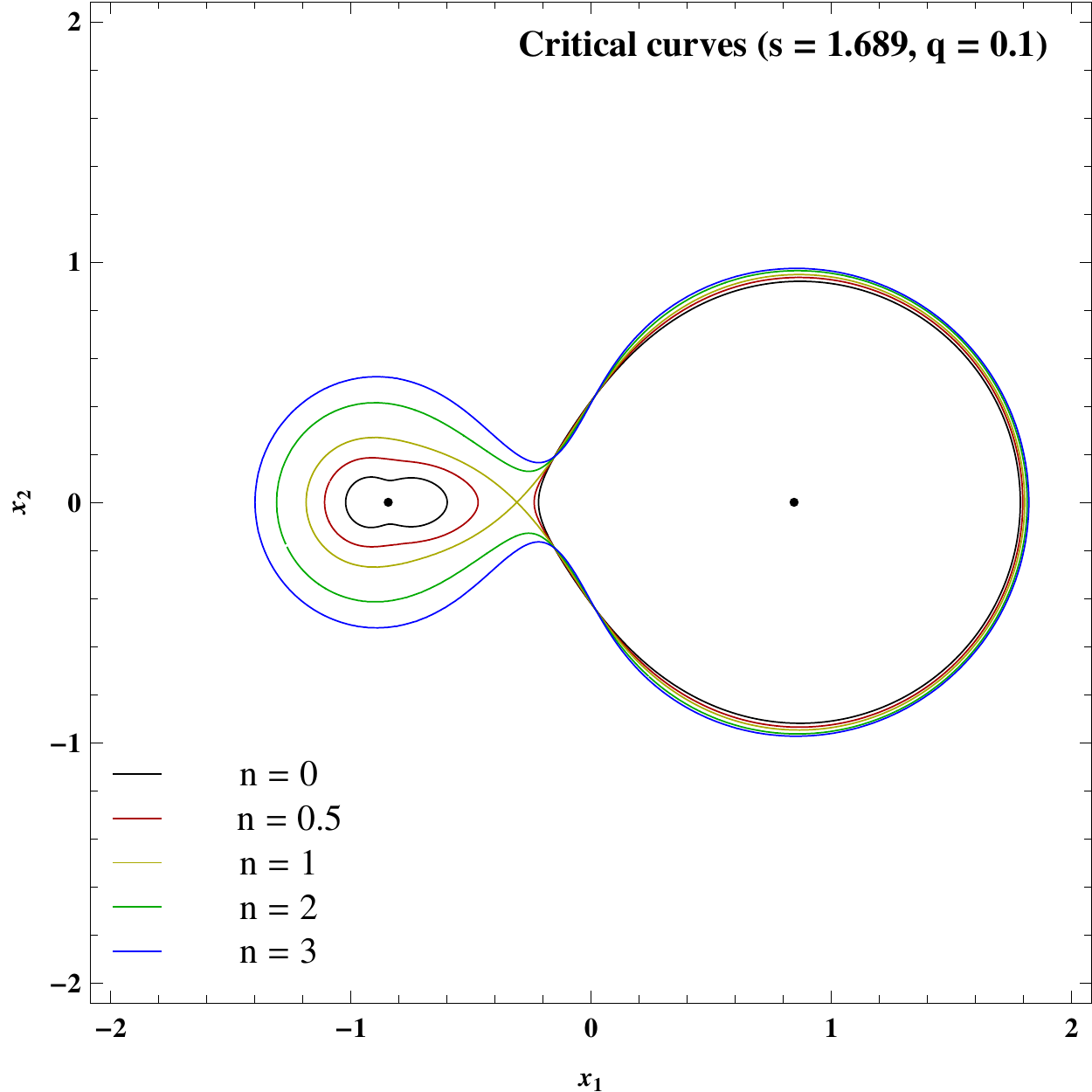}\hspace{4 mm}\includegraphics[width=6cm]{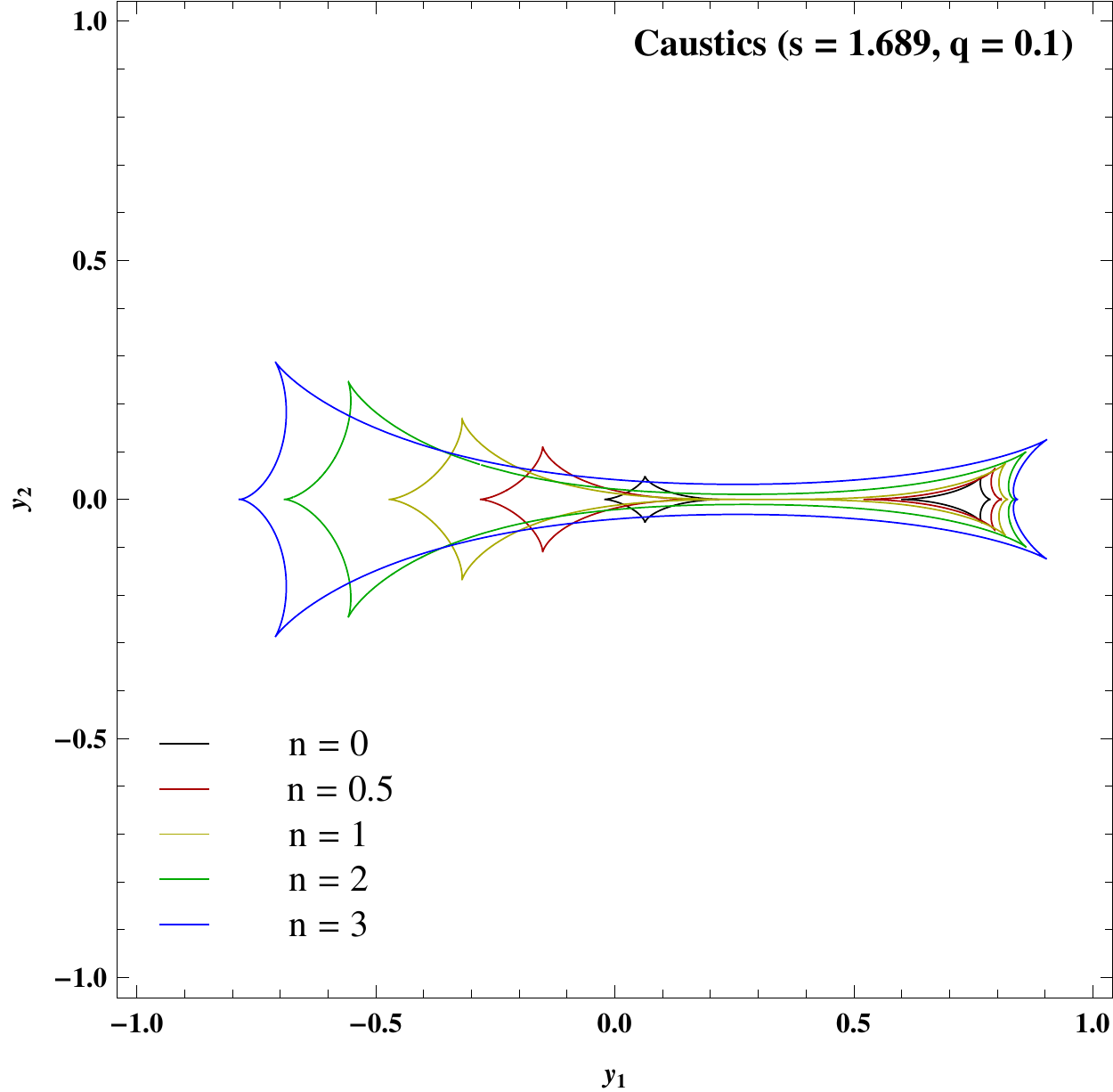}
\caption{Unequal-strength binary. Intermediate-Wide transition} \label{2iw}
\end{figure}

The transition between the intermediate and the wide regime occurs at $s=1.689$ for $n=1$ but is pushed to higher separations for $n>1$ and to lower separations for $n<1$. So, contrarily to what happens in the equal-strength case, also this transition is affected when we change $n$. In general, the intermediate regime occupies a larger volume in the parameter space for $n>1$ and a smaller volume for $n<1$.

\begin{figure}[t]
\centering
\includegraphics[width=6cm]{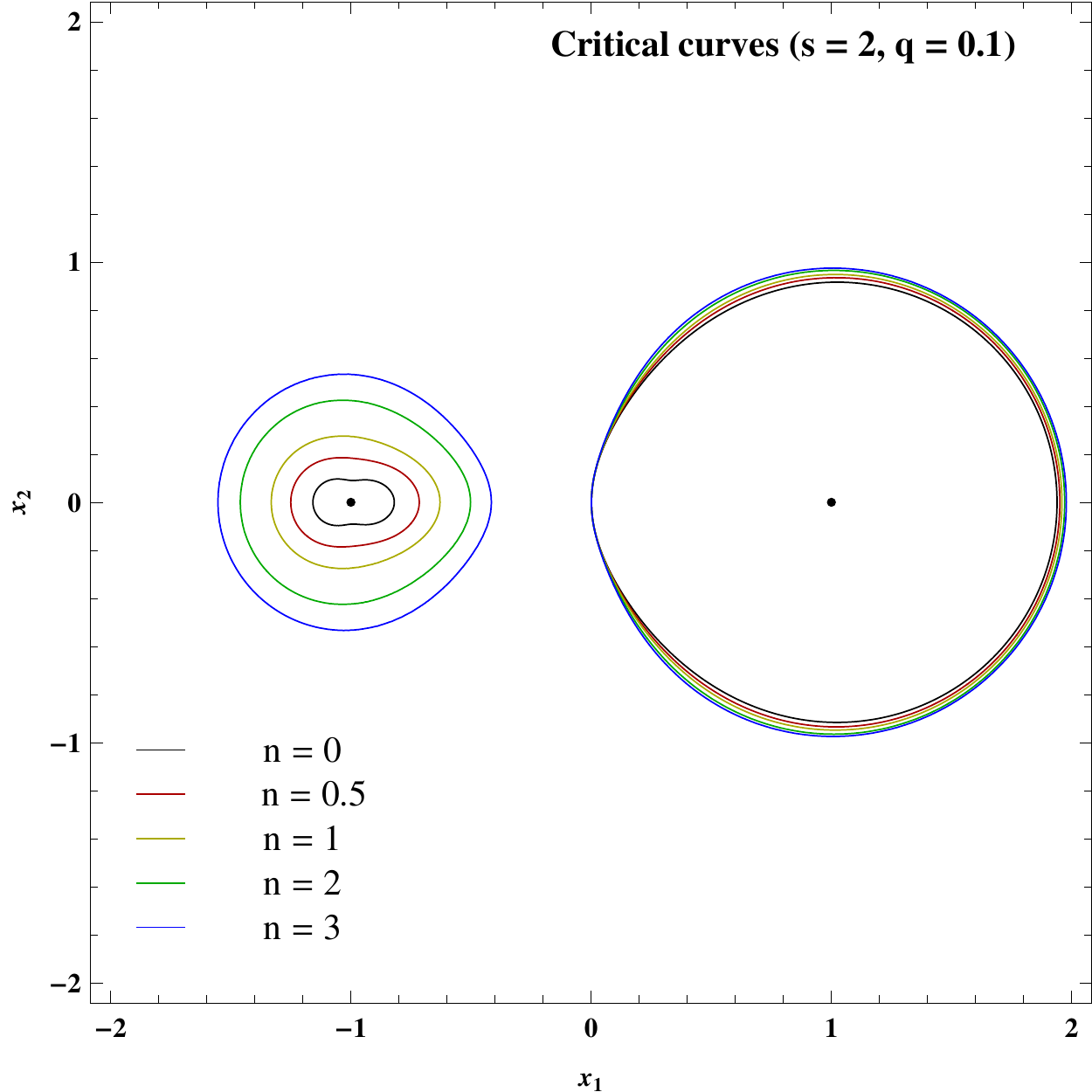}\hspace{4 mm}\includegraphics[width=6cm]{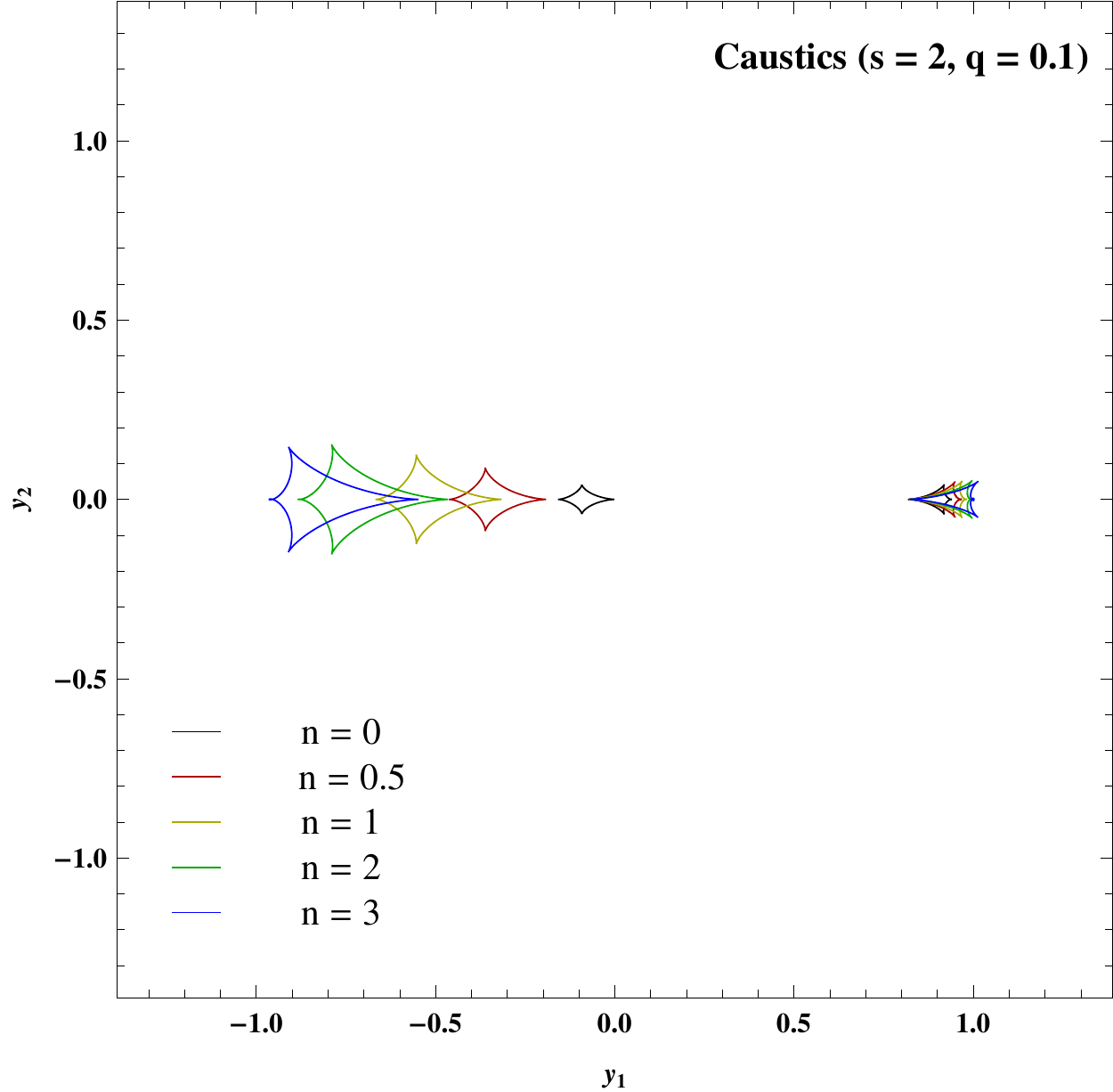}
\caption{Unequal-strength binary. Wide separation} \label{2w}
\end{figure}

Finally, Fig. \ref{2w} shows the wide regime $s=2$. Note that the weaker lens critical curve has a radius that strongly depends on $n$, since $\theta_E\sim \epsilon_i^{1/(n+1)}$. The caustic of the weaker lens, instead, remains of the same size and moves toward the projected lens position at higher $n$.

\subsection{Extreme unequal-strength binary $q=0.001$}

Now we go deeper into the unequal-strength regime of the binary lens, with $\epsilon_A\ll \epsilon_B$. In the standard $n=1$ context, this regime is conventionally called the ``planetary'' limit, as it naturally applies to the search for exoplanetary systems by the microlensing method. For $n=0$, it may represent gravitational lensing by a massive galaxy accompanied by a smaller satellite.

\begin{figure}[t]
\centering
\includegraphics[width=6cm]{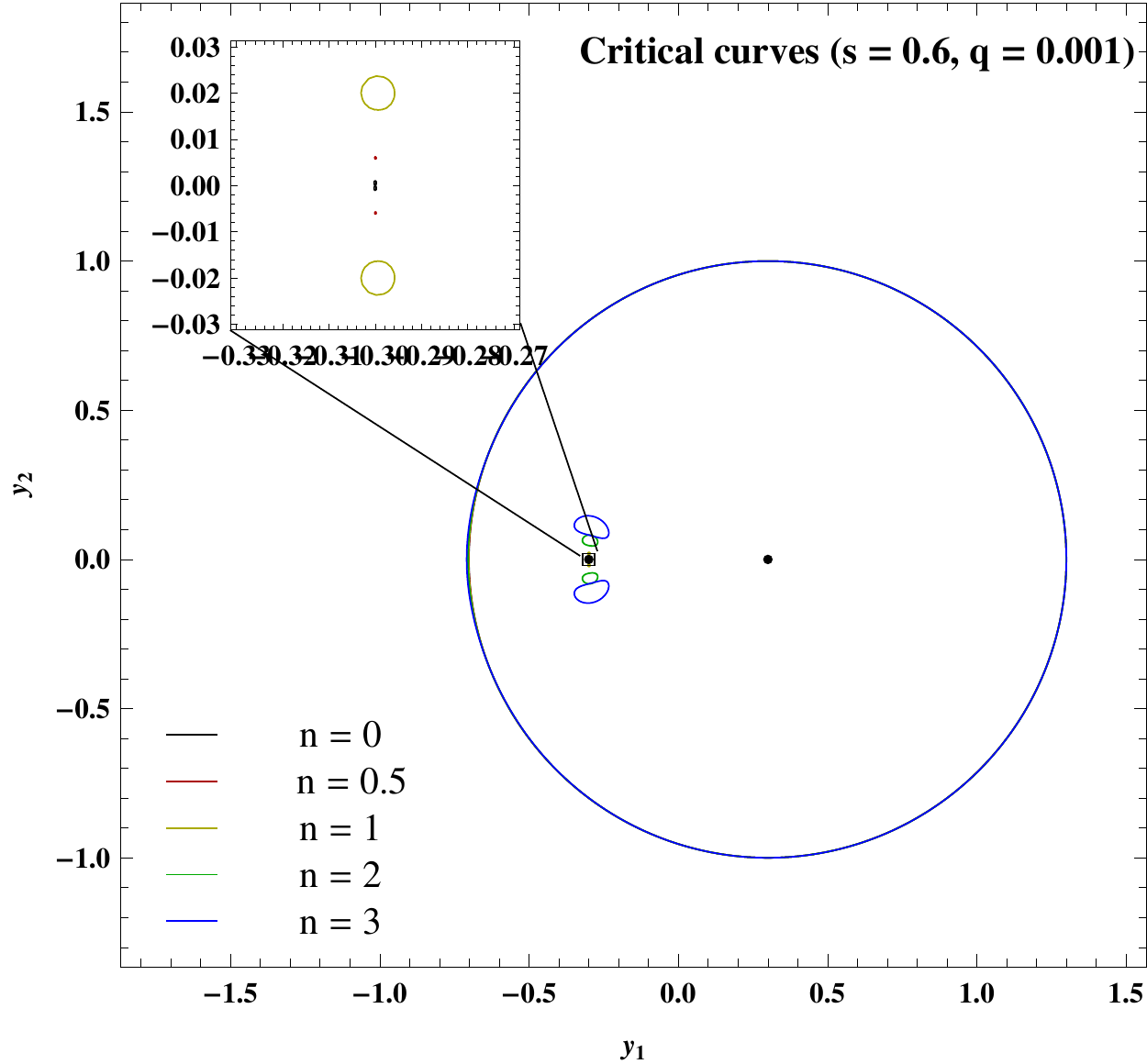}
\hspace{4 mm}
\includegraphics[width=6cm]{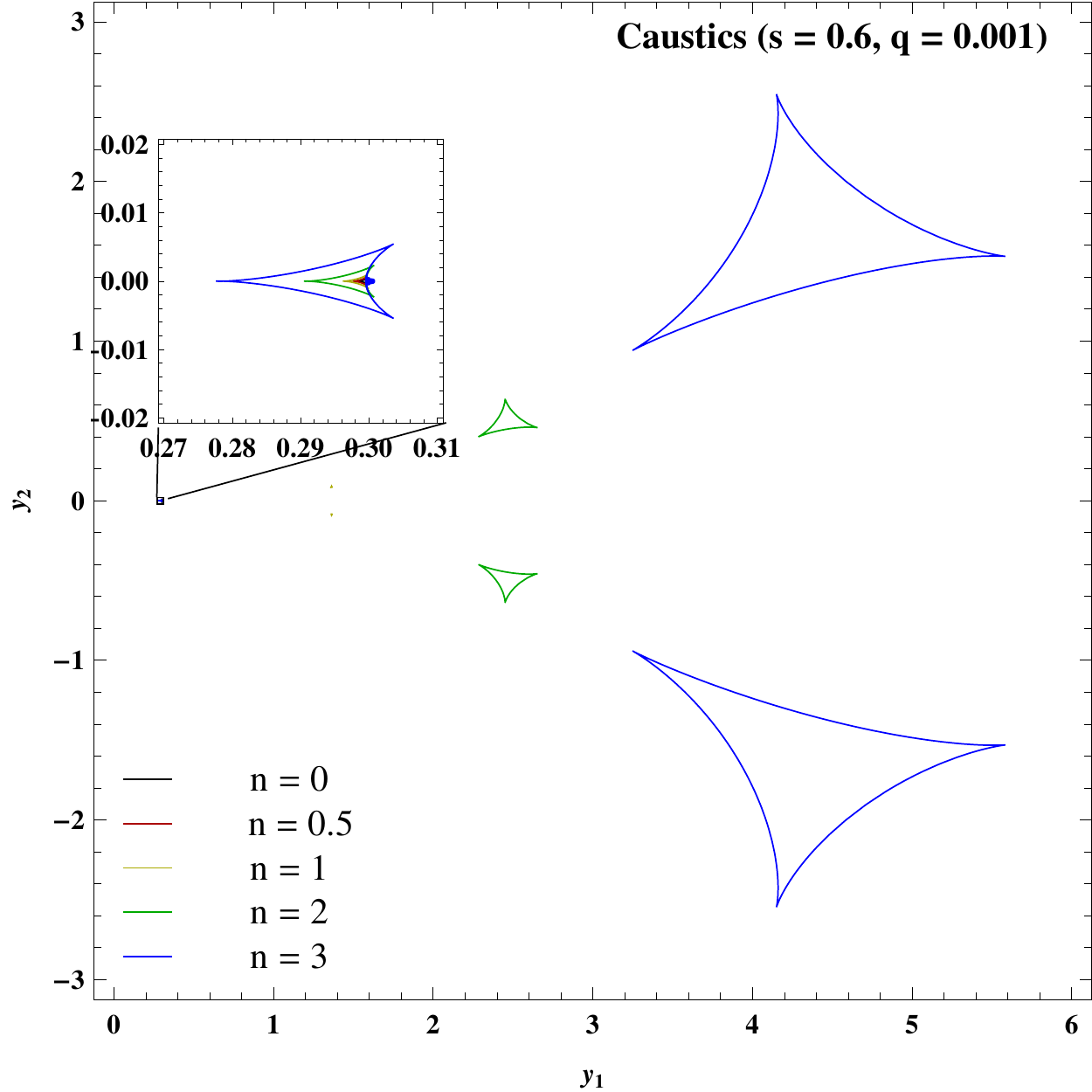}
\caption{Extreme unequal-strength binary. Close separation} \label{3c}
\end{figure}

Fig. \ref{3c} shows the close regime for $s=0.6$. The central caustic remains tiny, as the main critical curve is very weakly perturbed by the weaker lens. The small inner critical curves grow larger with $n$, as already noted in the previous subsection. The corresponding triangular caustics, become huge. This contrasts with the $n=1$ case, for which it is well known that the chance of detecting planets is related to the probability that the source trajectory crosses these very small caustics. In the $n>1$ case, it is much easier to detect a weaker component.

\begin{figure}[t]
\centering
\includegraphics[width=6cm]{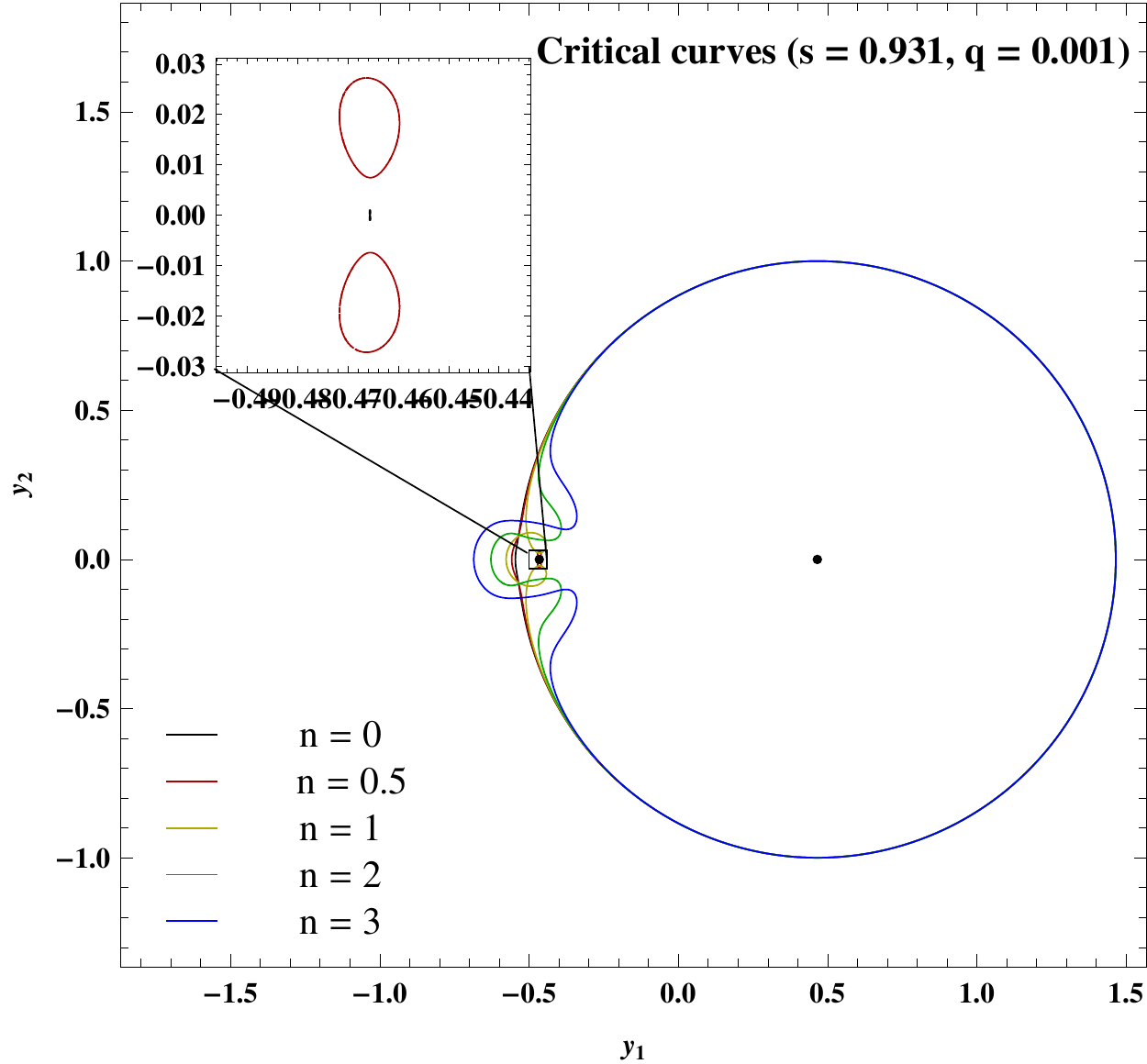}\hspace{4 mm}\includegraphics[width=6cm]{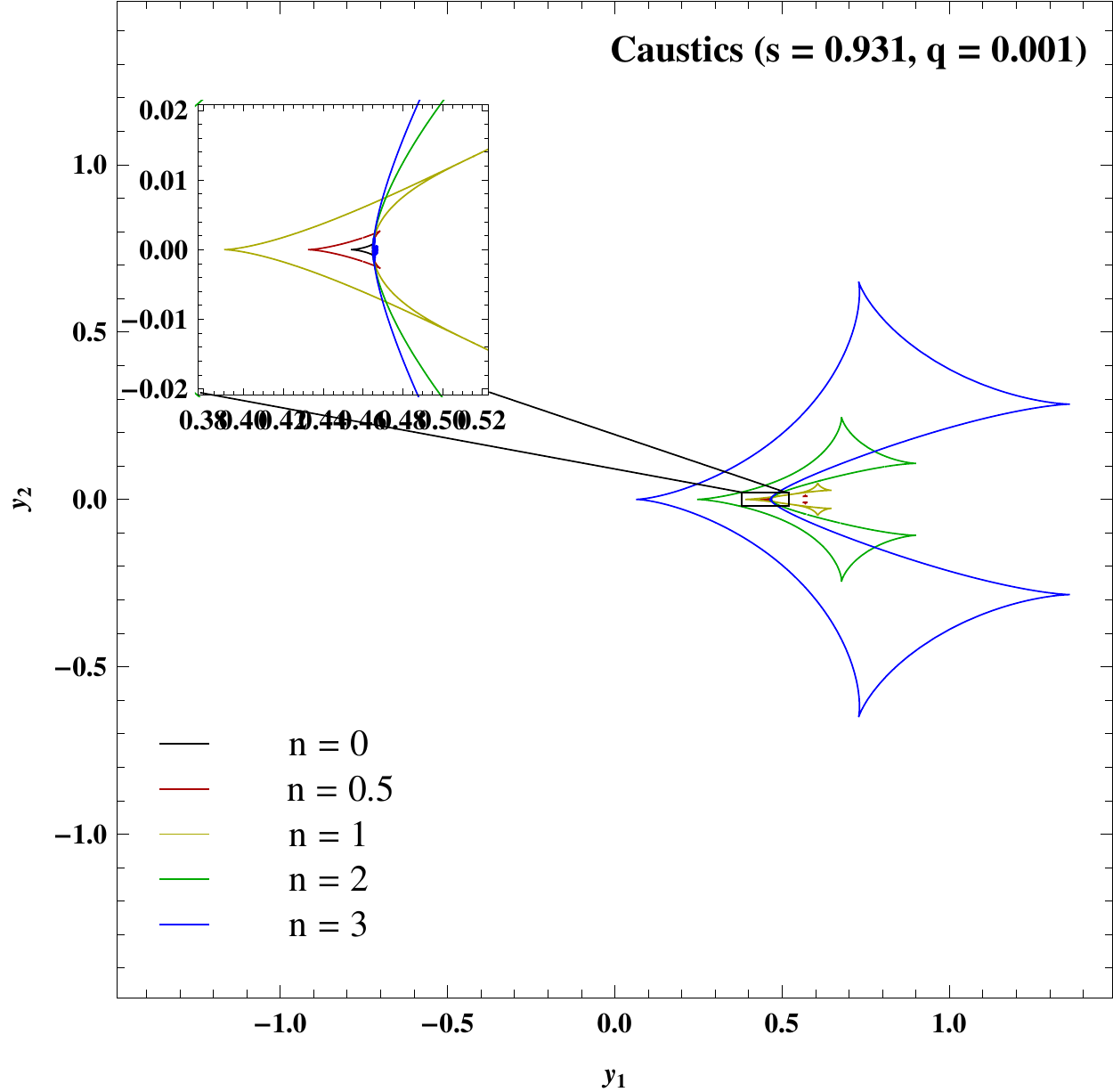}
\caption{Extreme unequal-strength binary. Close-Intermediate transition} \label{3ci}
\end{figure}

The transition between the close and intermediate regime occurs at $s=0.931$ in the $n=1$ case. For $n>1$ we are already in the intermediate regime here (Fig. \ref{3ci}), while we are still in the close regime for $n<1$, although the secondary critical curves and caustics are almost invisible.

\begin{figure}[t]
\centering
\includegraphics[width=6cm]{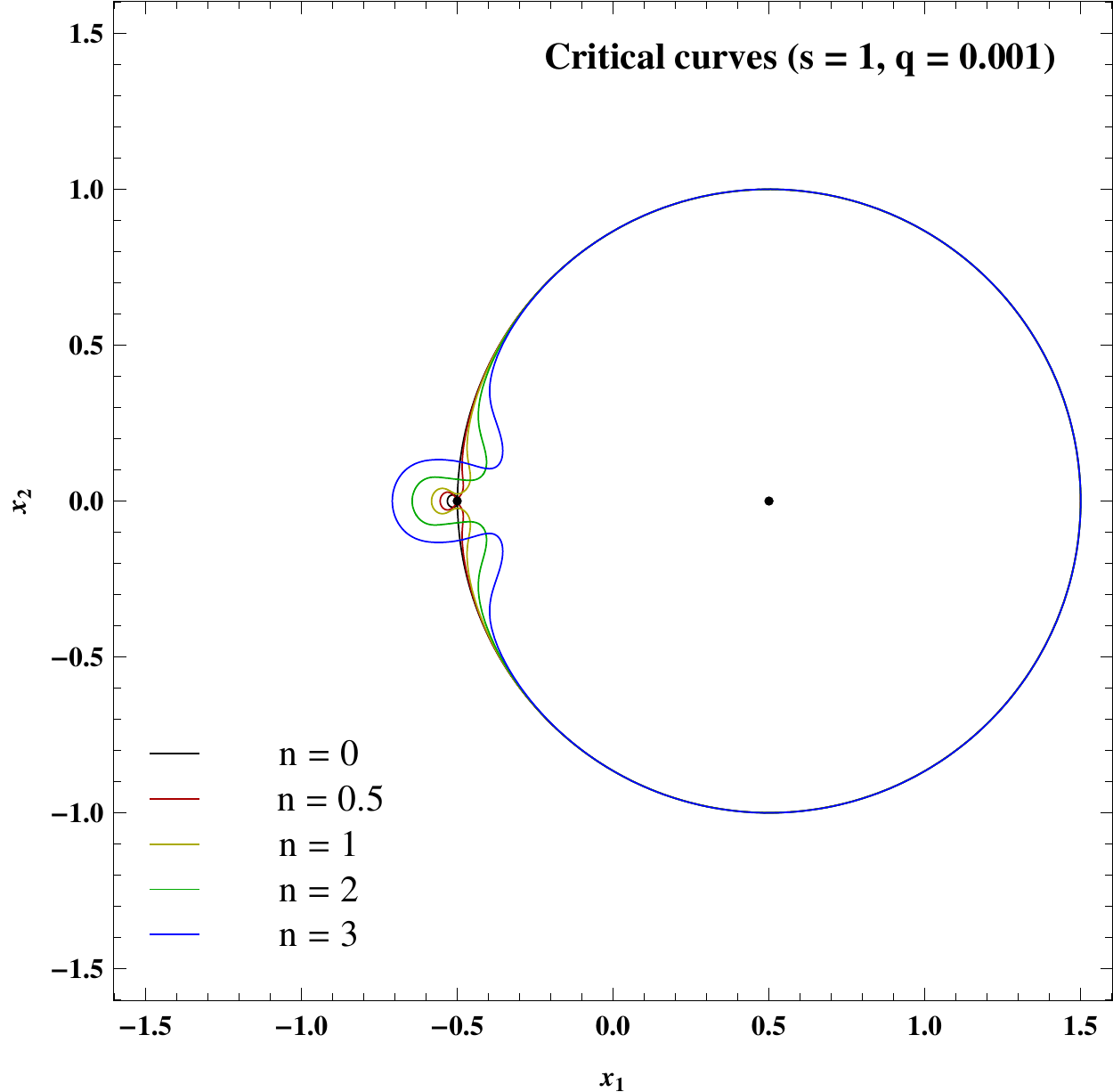}\hspace{4 mm}\includegraphics[width=6cm]{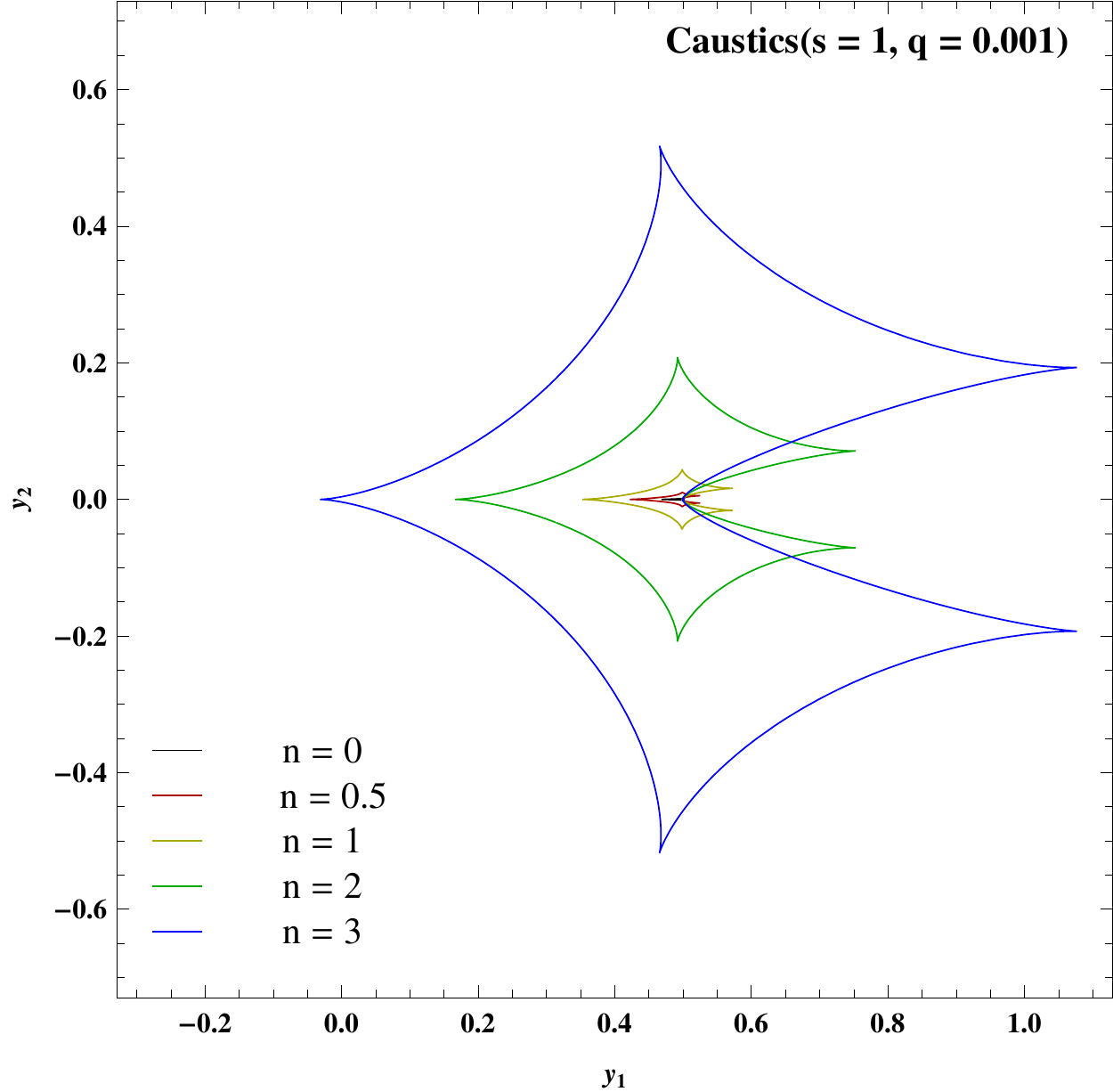}
\caption{Extreme unequal-strength binary. Intermediate separation} \label{3i}
\end{figure}

At $s=1$ (Fig. \ref{3i}), we are in the so-called resonant regime, for which the $n=1$ caustic assumes the largest size. Actually, for $n>1$ the total size of the caustics is similar to that in the close regime, which is dominated by the triangular caustics.

\begin{figure}[t]
\centering
\includegraphics[width=6cm]{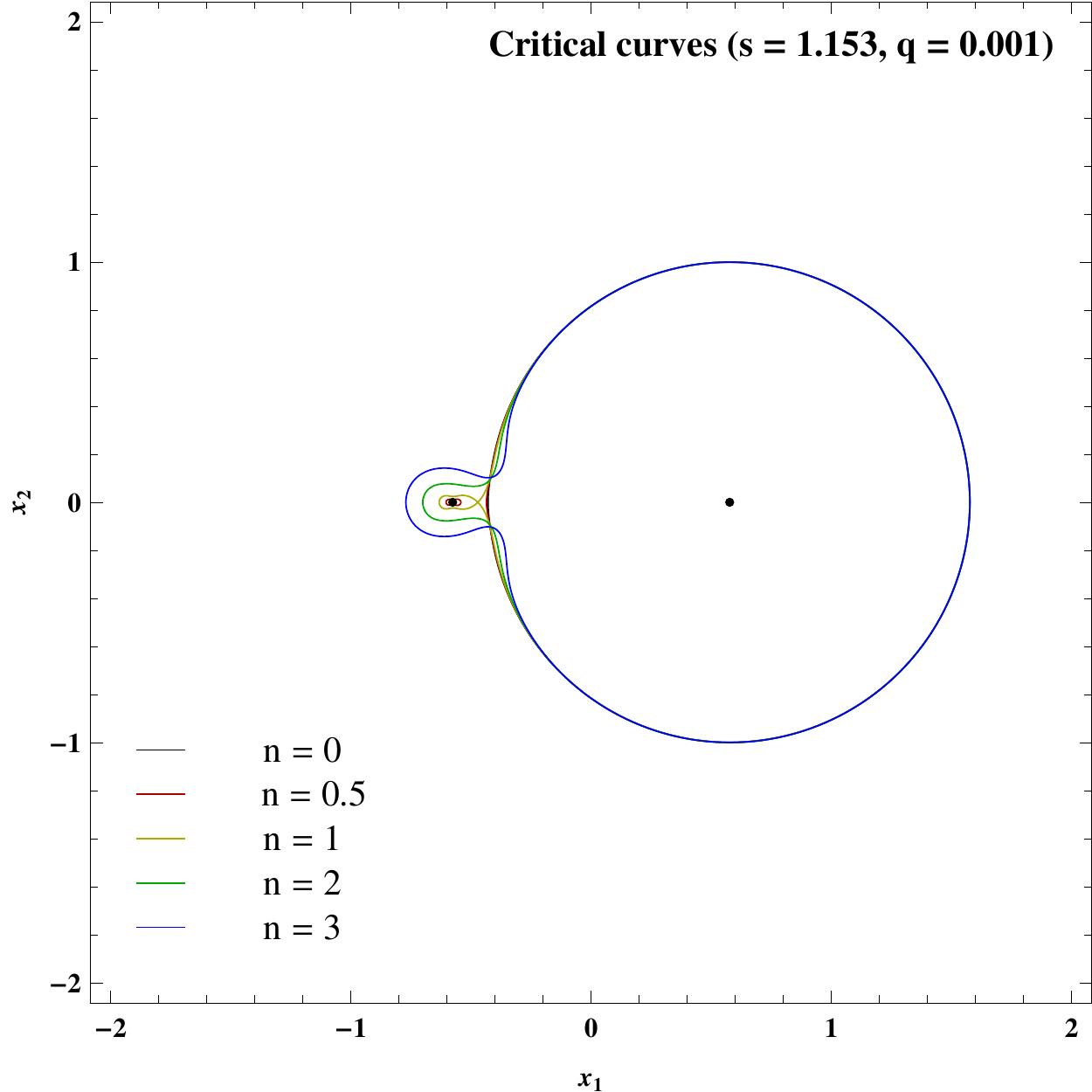}\hspace{4 mm}\includegraphics[width=6cm]{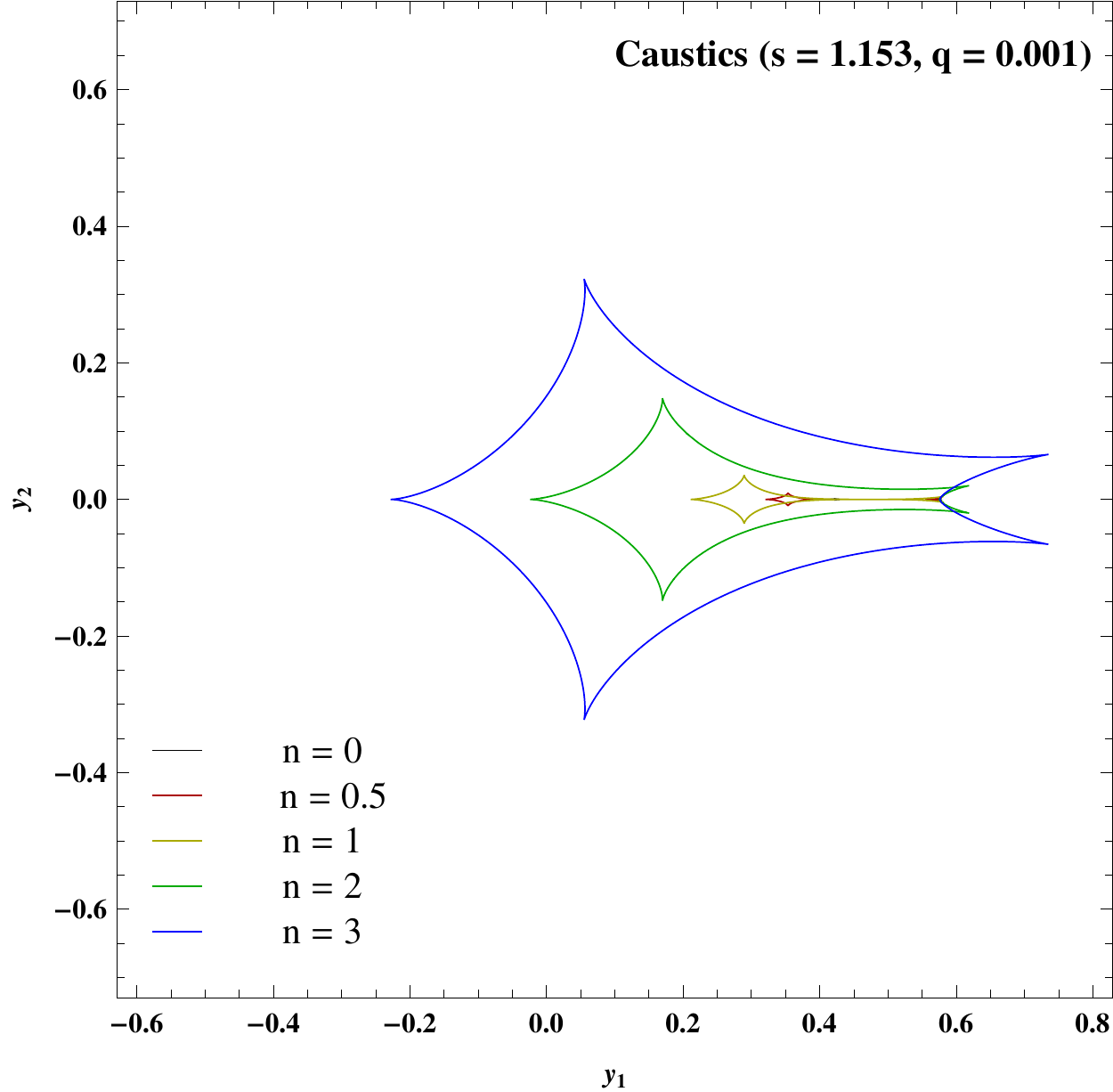}
\caption{Extreme unequal-strength binary. Intermediate-Wide transition} \label{3iw}
\end{figure}

At $s=1.153$, the $n=1$ caustic enters the wide regime, while the $n>1$ ones are still in the intermediate regime. The caustics rapidly evolve but still the $n>1$ ones remain larger (Fig. \ref{3iw}). In general, the $n<1$ caustics are extremely small.

\begin{figure}[t]
\centering
\includegraphics[width=6cm]{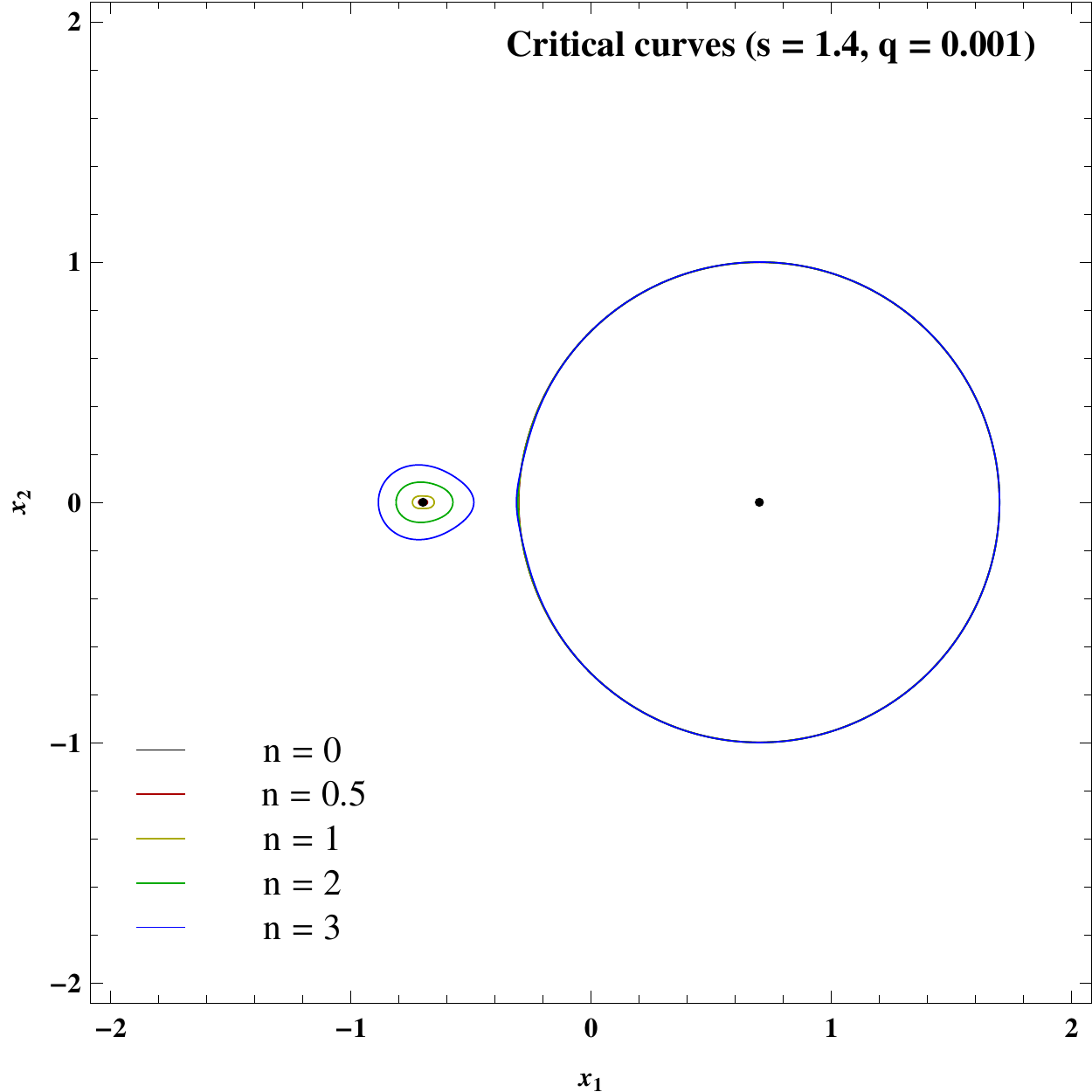}\hspace{4 mm}\includegraphics[width=6cm]{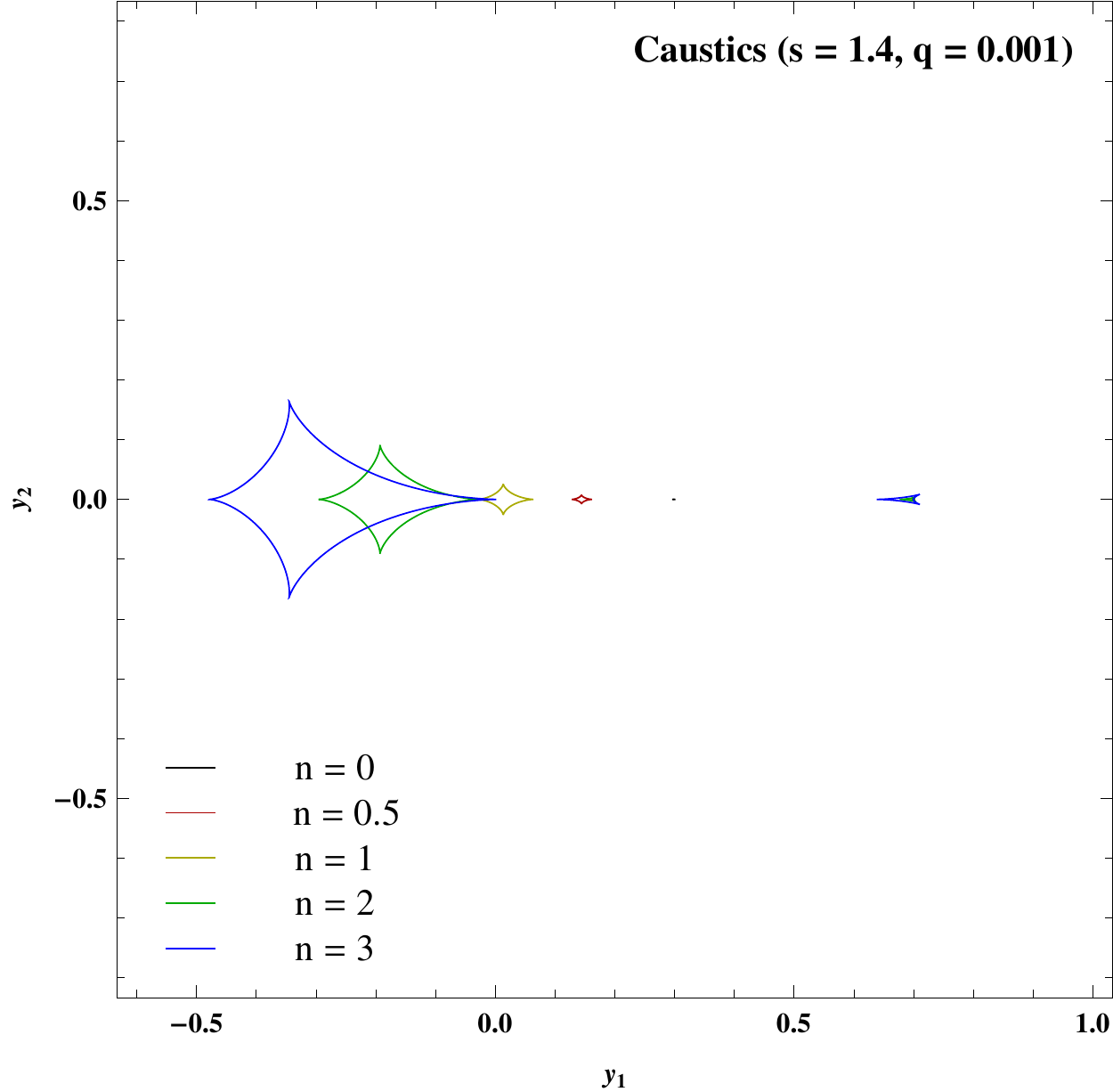}
\caption{Extreme unequal-strength binary. Wide separation} \label{3w}
\end{figure}

In the wide regime (Fig. \ref{3w}), we have two separate critical curves. The weaker lens still has a larger ring in the $n>1$ case, as discussed before. The caustics tend to become of similar size, as in the $q=0.1$ case, although at $s=1.4$ we still appreciate that the $n>1$ caustics are slightly larger than the $n=1$ one and are closer to the projected lens position.

\section{Boundaries of the three topology regimes}

In the standard Schwarzschild case (see Ref. \cite{ErdSch}), the transition between the close and intermediate regime occurs at a separation $s_{CI,n=1}$ implicitly given by the solution of the equation
\begin{equation}
\dfrac{q}{(q+1)^{2}}=\dfrac{(1-s_{CI,n=1})^{3}}{27 s_{CI,n=1}^{8}}, \label{sCI1}
\end{equation}
while the transition between the intermediate and wide regime falls at
\begin{equation}
s_{IW,n=1}=\dfrac{\left(q^{\frac{1}{3}}+1\right)^{\frac{3}{2}}}{(q+1)^{\frac{1}{2}} }. \label{sIW1}
\end{equation}

Our numerical exploration has shown that the three topologies known in the $n=1$ cases (close, intermediate, wide) naturally propagate to the $n>1$ case, with no change, save for the sizes and elongation of critical curves and caustics. However, we have already noted how the region in the parameter space corresponding to the intermediate regime widens with respect to the standard case, with the only exception of the equal-strength case, for which the intermediate-wide transition still occurs for $s=2$ for any $n$. In the $n<1$ case, instead, the region corresponding to the intermediate regime shrinks and we have an additional catastrophe (the elliptic umbilic) well within the close regime. However, this catastrophe does not change the overall caustic topology.

In this section we aim at finding the boundaries of the three topology regimes in the parameter space for any $n$ and the position of the elliptic umbilic for $n<1$.

Transitions occur via higher order singularities of the lens map, namely beak-to-beak singularities in the binary lens case. So, in order to find the values of the parameters $q$ and $s$ for which such transitions occur, we need to impose that the Jacobian $J$ and its derivative $\partial J/\partial z$ simultaneously vanish. This quantity reads

\begin{eqnarray}
\dfrac{\partial J}{\partial z}&=\dfrac{1-n^{2}}{4} \bigg[ \dfrac{\epsilon_{A}}{(z+s/2)^{\frac{n+3}{2}}(\bar z +s/2) ^{\frac{n+1}{2}}}+\dfrac{\epsilon_{B}}{(z-s/2)^{\frac{n+3}{2}}(\bar z -s/2) ^{\frac{n+1}{2}}} \bigg]\nonumber \\ &\bigg\lbrace 2+ (n-1)\bigg[ \dfrac{\epsilon_{A}}{(z+s/2)^{\frac{n+1}{2}}(\bar z +s/2) ^{\frac{n+1}{2}}}+\dfrac{\epsilon_{B}}{(z-s/2)^{\frac{n+1}{2}}(\bar z -s/2) ^{\frac{n+1}{2}}} \bigg]  \bigg\rbrace \nonumber \\
& \quad -\dfrac{(n-1)(n+1)^{2}}{8}\left\vert  \dfrac{\epsilon_{A}}{(z+s/2)^{\frac{n+1}{2}}(\bar z +s/2) ^{\frac{n+3}{2}}}+\dfrac{\epsilon_{B}}{(z-s/2)^{\frac{n+1}{2}}(\bar z -s/2) ^{\frac{n+3}{2}}} \right\vert^{2}\nonumber \\ & \quad +\dfrac{(n+3)(n-1)^{2}}{8} \bigg \lbrace \bigg[  \dfrac{\epsilon_{A}}{(z+s/2)^{\frac{n+5}{2}}(\bar z +s/2) ^{\frac{n-1}{2}}}+\dfrac{\epsilon_{B}}{(z-s/2)^{\frac{n+5}{2}}(\bar z -s/2) ^{\frac{n-1}{2}}}  \bigg]\nonumber \\ &\bigg[\dfrac{\epsilon_{A}}{(z+s/2)^{\frac{n-1}{2}}(\bar z +s/2) ^{\frac{n+3}{2}}}+\dfrac{\epsilon_{B}}{(z-s/2)^{\frac{n-1}{2}}(\bar z -s/2) ^{\frac{n+3}{2}}} \bigg]\bigg\rbrace,
\end{eqnarray}
which looks extremely complicated. However, we have managed to find the simultaneous solution of the equations $\partial J/\partial z=0$ and $J=0$ corresponding to the intermediate-wide transition. Unfortunately, after many efforts, we have not been able to find the  close-intermediate transition.

Concerning the intermediate-wide transition, we know that the beak-to-beak singularity occurs along the line joining the two lenses. This can be expressed by requiring $z=\bar z$ in complex notations. Then we introduce a new variable
\begin{equation}
y=\dfrac{s/2-z}{s/2+z},
\end{equation}
replacing $z$ in J and $ \partial J /\partial z $.
We then solve $ \partial J /\partial z = 0 $ in terms of $ y $, obtaining
\begin{equation}
y=\left(\dfrac{\epsilon_{B}}{\epsilon_{A}} \right)^{\frac{1}{2+n}}
\end{equation}
Finally, we insert this expression for $ y $ in $ J $ and solve $ J=0 $ for $ s $. In this way, we obtain:
\begin{equation}
s_{WI}=(q+1)^{-\frac{1}{n+1}} \left(q^{\frac{1}{n+2}}+1\right)^{\frac{n+2}{n+1}}
\end{equation}
which  reduces to the solutions already known (Eq. \ref{sIW1}) for $ n=1 $ \cite{ErdSch} and for the $n=0$ case \cite{ShinEvans}.

\begin{figure}[htbp]
\centering
\includegraphics[width=6cm]{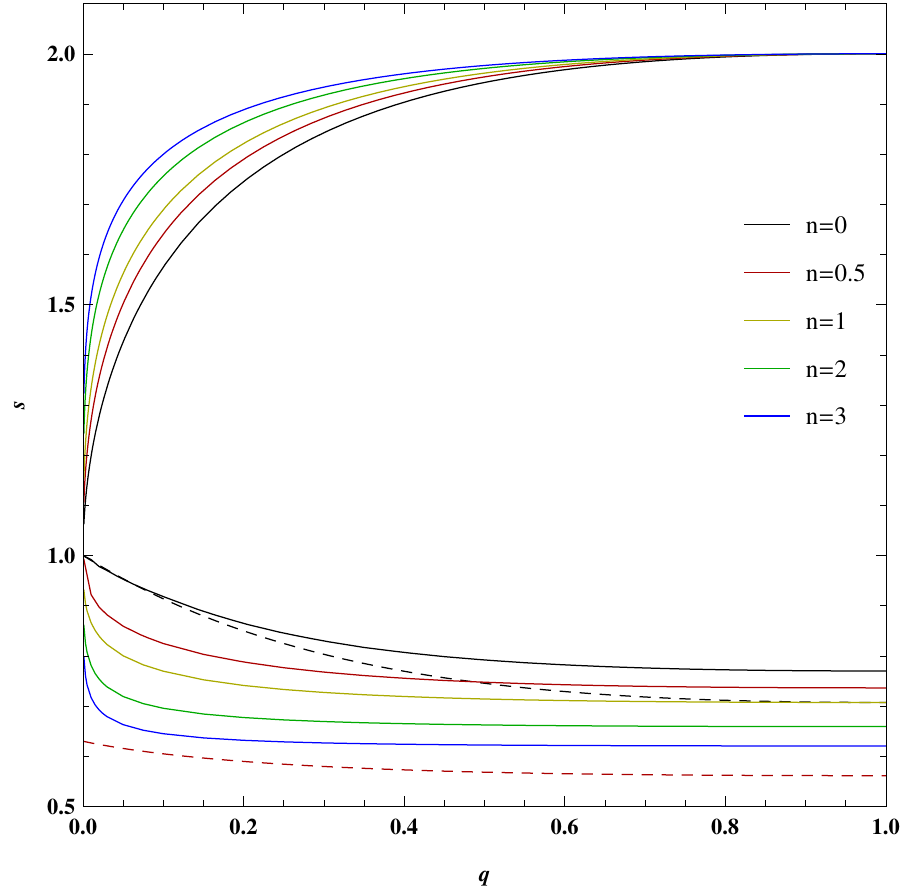}
\caption{The critical values of the projected separation (in units of $ \theta_{E} $) at which the caustics topology of a binary lens changes as a function of the mass ratio $  q $ and the index $ n $. The upper curves is $ s_{WI} $, i.e. the critical value of $ s $ between the wide caustic and the intermediate or resonant caustic topology. The lower curves correspond to $ s_{IC} $, the critical value between the resonant caustic and close caustic topology.}\label{6t}
\end{figure}

For the  Close-Intermediate transition, we are unable to extend Eq. (\ref{sCI1}) to arbitrary values of $n$, but we can easily study the transition numerically. Fig \ref{6t} plots $ s_{IC} $ and $ s_{WI} $ as functions of $ q $ for $  n=(0,0.5,1,2,3) $.

Note that for equal-mass binaries, $ q=1 $ the critical value of the intermediate-wide separation is always 2 for any index $n$, as already anticipated in the previous section.

In addition to the main topology transitions, we have discussed the existence of an elliptic umbilic catastrophe in the close regime for $n<1$. Shin and Evans \cite{ShinEvans} showed that this catastrophe should lie on a circle centered half-way between the two lenses and passing through them. We find that their argument is still valid for any values of $n$ and therefore look for simultaneous solutions of the equations $J=0$ and $\partial J/\partial z=0$ along this circle. After setting $z=s e^{i\theta}/2$, we introduce the variable $t$ by the equation
\begin{equation}
t^{\frac{1}{1+n}}\equiv \tan \frac{\theta}{2}.
\end{equation}
This considerably simplifies the two equations and we obtain the angular position of the elliptic umbilic as
\begin{equation}
t=\frac{\epsilon_B}{\epsilon_A},
\end{equation}
and the critical value of the separation at which the catastrophe occurs
\begin{equation}
s=(1-n)^{\frac{1}{1+n}} \sqrt{\epsilon_A^{\frac{2}{1+n}}+\epsilon_B^{\frac{2}{1+n}}}, \label{Equmb}
\end{equation}
which nicely matches the $n=0$ limit already found in Ref. \cite{ShinEvans}. As shown by this formula, for $n\rightarrow 1$ the elliptic umbilic occurs at zero separation and is therefore unobservable from $n=1$ on. In Fig. \ref{6t} the positions of the elliptic umbilic in the parameter space are shown by dashed curves in the cases $n=0$ and $n=0.5$.

\section{Analytical approximation of Critical curves and Caustics}

Our numerical study of the critical curves and caustics of exotic lenses has shown some interesting differences with respect to the standard Schwarzschild case. In particular, the enormous size of the formerly small triangular caustics in the close topology represents the most impressive novelty of the systems explored in our investigation.

These numerical results call for support by analytical formulae describing the behavior of the caustics in some particular limits. In the binary Schwarzschild lens, this kind of exploration has led to several interesting analytical results in the close, wide and planetary limits \cite{Dom,Boz1,Boz2,Boz3,An,Han,Chung}. Some analytical results have also been obtained for the $n=0$ case \cite{ShinEvans}.
In this section we extend these studies to a generic value of the exponent $n$. By discussing the analytical approximations, we will gain a much deeper understanding of the evolution of the caustic structure with $n$.

\subsection{Wide binary}
First, let us consider the simplest limit: an isolated object whose deflecting potential is perturbed by another object at a distance much larger than the Einstein radius.

We place our lens at position $ z_{A}=0 $, and the perturbing object at $ z_{B}=-s $, with $s\gg 1$. So the lens equation reads:
\begin{equation}
\zeta=z-\frac{\epsilon_{A}}{z^{\frac{n-1}{2}}\bar{z}^{\frac{n+1}{2}}} -\frac{\epsilon_{B}}{(z+s)^{\frac{n-1}{2}}(\bar{z}+s)^{\frac{n+1}{2}}} \label{LEwide}
\end{equation}
Their properties can be studied through the Jacobian determinant of the lens map
\begin{eqnarray}
J &=\left\lbrace 1 + \dfrac{n-1}{2}\left[ \dfrac{\epsilon_{A}}{z^{\frac{n+1}{2}}\bar{z} ^{\frac{n+1}{2}}}+\dfrac{\epsilon_{B}}{(z+s)^{\frac{n+1}{2}}(\bar{z}+s) ^{\frac{n+1}{2}}} \right] \right\rbrace ^{2} \nonumber \\
& \quad -\dfrac{(n+1)^{2}}{4}\left\vert  \dfrac{\epsilon_{A}}{z^{\frac{n+3}{2}}\bar{z} ^{\frac{n-1}{2}}}+\dfrac{\epsilon_{B}}{(z+s)^{\frac{n+3}{2}}(\bar{z}+s) ^{\frac{n-1}{2}}} \right\vert^{2}. \label{Jacwide}
\end{eqnarray}
Starting from an isolated point-lens, the presence of the other mass, that is very far from the first one, deforms the circular critical curve. This deformation can be recovered by a perturbative approach \cite{Dom,Boz2}. Therefore, we set
\begin{equation}
z=\rho(1+ \delta) e^{i\theta},
\end{equation}
with $\delta \sim 1/s^p$, and the exponent $p$ being fixed by the perturbative expansion.

At zero order, sending $s$ to infinity in Eq. (\ref{Jacwide}), the equation $J=0$ is solved by
\begin{equation}
\rho=\epsilon_{A}^{1/(n+1)},
\end{equation}
which is the critical curve for a single lens.

To first order in $1/s$, the first correction to the critical curve appears for  $\delta \sim 1/s^{n+1}$:
\begin{equation}
\delta=\dfrac{\epsilon_{B}[1-n + (n+1)\cos(2\theta) ]}{2(n+1)s^{n+1}}.
\end{equation}
In the standard $ n=1$ case, this correction reduces to
\begin{equation}
\delta=\dfrac{\epsilon_{B}\cos(2\theta) }{2s^{2}},
\end{equation}
which matches what was found in previous studies \cite{Dom,Boz2}.
The deformation of the circular critical curve is still given by $\cos(2\theta)$ with the same amplitude independent of $n$. However, the curve is slightly shrunk (or enlarged) by a fixed term proportional to $(1-n)/(n+1)$.

By applying the lens equation (\ref{LEwide}) and expanding we obtain:
\begin{eqnarray}
 Re[\zeta(\theta)]&=& -\dfrac{\epsilon_{B}}{s^{n}}+ \dfrac{(n+1)\epsilon_{A}^{1/(n+1)}\epsilon_{B}\cos^{3}\theta}{s^{n+1}}\\
 Im[\zeta(\theta)]&=&-\dfrac{(n+1)\epsilon_{A}^{1/(n+1)}\epsilon_{B} \sin^{3}\theta}{  s^{n+1}}
\end{eqnarray}
For $ \epsilon_{B}\rightarrow 0 $ we return to the case of a point-lens, therefore the terms with $ \epsilon_{B} $ give the perturbation of the caustics.

The first term in $ Re[\zeta] $ is a shift $ -\epsilon_{B} /s^{n}$ of the caustic in the direction toward the perturbing object. The shift decays with the index $n$, in agreement with what we note in Figs. \ref{1w}, \ref{2w}, \ref{3w}.

The other terms in  $ \zeta $ describe the shape of the caustics as $ \cos^{3}\theta+i \sin^{3}\theta$, which corresponds to the classic Chang-Refsdal 4-cusped astroid \cite{ChaRef,Boz2}. This does not change from $n=1$ to greater or smaller $n$. However, the size of the caustic scales as $(n+1)/s^{n+1}$. At moderate separations, the caustics at higher $n$ are larger than in the $n=1$ case, but in the deep wide limit $s \gg 1$, the caustics become smaller and smaller.

\subsection{Close binary}

The close binary regime poses the greatest challenge for any analytical results, as we already experienced with our search for the boundary between the close and intermediate regime. We will first discuss the central caustic \cite{Dom,Boz2} and then the ``secondary'' caustics \cite{Boz3}, which for $n>1$ give rise to the huge triangular structures shown in our numerical investigation.

\subsubsection{Central caustic}

We leave the two masses at positions $ z_{A}=-s/2 $ and $ z_{B}=s/2$ as in the lens equation (\ref{LensEq}) and consider the Jacobian determinant in the form of Eq. (\ref{Jac}).

The separation between the masses $s$ modulate the deviations from the Schwarzschild lens and constitute the perturbative parameter in our expansion. We use the parametrization
\begin{equation}
z=\rho(1+ \delta_1+\delta_2) e^{i\theta},
\end{equation}
where $ \delta_{1} $ is of order $ s $, and $ \delta_{2} $ is of order $ s^{2} $ . Substituting in equation (\ref{Jac}) and expanding in powers of $ s $ up to second order, we can solve for $ \rho$, $\delta_{1} $ and $\delta_{2}$ :
\begin{align}
\rho& = 1\nonumber \\
\delta_{1}& =\frac{1}{2}(-\epsilon_{A}+\epsilon_{B})\cos\theta \ s  \nonumber \\
\delta_{2}& = \frac{1}{16} \lbrace   - 1+ 4n \epsilon_{A}\epsilon_{B} + [ 1+(16+4n) \epsilon_{A}\epsilon_{B} ] \cos (2 \theta) \rbrace  \ s^{2}
\end{align}
Note that, the dependence on $ n $ is only in the second order of the expansion. Once known the perturbations to the critical curve, we can find the caustic using the lens equation:
\begin{eqnarray}
 Re[\zeta(\theta)]&=&\dfrac{(-\epsilon_{A}+\epsilon_{B})}{2(\epsilon_{A}+\epsilon_{B})}s+ \dfrac{(n+1)\epsilon_{A}\epsilon_{B}\cos^{3}\theta}{(\epsilon_{A}+\epsilon_{B})^{\frac{2n+3}{n+1}}}s^{2}\\
 Im[\zeta(\theta)]&=&-\dfrac{(n+1)\epsilon_{A}\epsilon_{B}\sin^{3}\theta}{(\epsilon_{A}+\epsilon_{B})^{\frac{2n+3}{n+1}}}s^{2}
\end{eqnarray}
The first order perturbations in $Re[\zeta] $ is simply a shift term that displaces the caustic from the origin (median point between the two lenses) to the center of strength (the wording ``center of mass'' should be reserved to $n=1$ case only).

The dependence on $ n $ appears in the second order terms in $ \zeta$ with the factor $ (n+1) $. The denominator is just a power of $(\epsilon_A+\epsilon_B)$, which has been set to 1 in our investigations. Then we learn that the central caustic scales linearly with the exponent $n$, something that can be appreciated in Fig. \ref{3c}, in particular. The same occurs for all values of the strength ratio $q$. The shape of the caustic is still the classical 4-cusped astroid as in the wide separation limit, reminding us of the wide-close degeneracy \cite{Dom}, which plagues gravitational lensing by binary systems.

\subsubsection{Secondary caustics}

The lens map is characterized by secondary critical curves that can appear in the form of small ovals near the center of strength of the system (Fig. \ref{1c}) \cite{Boz2,Boz3}. These secondary critical curves produce tiny caustics for $n<1$ and the huge triangular caustics far from the lens system for $n>1$. So, we are particularly interested in confirming our numerical results by an analytical approximation that makes us understand how this phenomenology arises.

Our starting point is still the Jacobian in the form Eq. (\ref{Jac}). The critical curves appear along a circle with radius $s/2$ centered in our origin. Unfortunately, we are unable to achieve a full description of these small ovals and we have to content ourselves with some partial yet valuable information.

By setting
\begin{equation}
z=\frac{s}{2} e^{i\theta},
\end{equation}
we manage to obtain the two intersections of these ovals with the circle of radius $s/2$. These two points are already enough to appreciate the size of the ovals and then of the corresponding tiny (for $n\leq 1$) or giant (for $n>1$) caustics. Inserting this parametrization in Eq. (\ref{Jac}) and saving the zero order in $s$ only, we obtain an equation for $\theta$:
\begin{equation}
-\text{$\epsilon_{A} $} \text{$\epsilon_{B} $} e^{2 \imath n \pi  } (n^2+1) \sin \theta \sin ^n(\theta/2) \cos ^n(\theta/2)+2n[ \text{$\epsilon_{A} $}^2  \sin ^{2 n+2}(\theta/2)+\text{$\epsilon_{B} $}^2  \cos ^{2 n+2}(\theta/2)]=0.
\end{equation}

This equation is satisfied by the angles:
\begin{eqnarray}
\theta_{1}=2 \tan ^{-1}\bigg[\left(\frac{\text{$\epsilon_{B} $}}{n\text{$\epsilon_{A} $} }\right)^{\frac{1}{n+1}}\bigg] \\
\theta_{2}=2 \tan ^{-1}\bigg[\left(\frac{n\text{$\epsilon_{B} $} }{\text{$\epsilon_{A}$}}\right)^{\frac{1}{n+1}}\bigg].
\end{eqnarray}

These two angles represent the positions of the two intersections of the secondary critical curves with the circle of radius $s/2$. Note that the standard case $n=1$ is degenerate, in the sense that the two intersections coincide at zero order. This is a warning that only in the special Schwarzschild lens the secondary critical curves require an expansion to the next order in $s$ of the Jacobian \cite{Boz3}. This is the root of the extreme difference in size we find when we move from $n=1$ to $n>1$. At $n>1$ the secondary critical curves open up at first order already, while they remain pointlike at $n=1$ and open up at second order only. Having widened our horizons by our exploration of all values of $n$, we learn that this fact is related to the presence of the elliptic umbilic at $n<1$, which collapses to $s=0$ for $n\rightarrow 1$. Therefore, in the starting point of our expansion at small $s$, the $n=1$ case is right at the elliptic umbilic, which makes secondary critical curves shrink to zero size.

Setting $ \epsilon_{A}=\epsilon_{B} $ in $ \theta_{1} $ and $ \theta_{2} $, we can easily see that $\theta_2-\pi/2=\pi/2-\theta_1$, i.e. the two intersections are symmetrical with respect to the vertical axis, as it is expected in the equal-strength case.

Using the lens equation, we can calculate the position of the caustic points corresponding to these intersections. We find

\begin{equation}
\zeta=\dfrac{(n^{-\frac{2}{1+n}}+1)^{\frac{n-1}{2}}}{2 n^{\frac{1}{1+n}}s^{n}}\left[\left(n^{\frac{n}{1+n}}-n^{\frac{1}{1+n}}\right) - i(n+1) \right] \label{causec}
\end{equation}

The imaginary part of these caustic points gives us an idea of the distance of the secondary caustic from the origin. The real part, instead, can be taken as a measure of the transverse extension of the caustic. Note that the latter is exactly zero in the limit $n=1$, which means that the extension of the caustic must be calculated at the next order in the power expansion in $s$ in the standard case. Indeed, from previous studies, \cite{Boz3}
we know that the transverse extension of the secondary caustic is proportional to $s^3$ for the standard binary Schwarzschild lens. When $n>1$, the secondary caustic has an extension comparable to its shift along the vertical axis, which is what we have observed in Figs. \ref{1c}, \ref{2c} and \ref{3c}.

\begin{figure}[h]
\centering
\includegraphics[width=8cm]{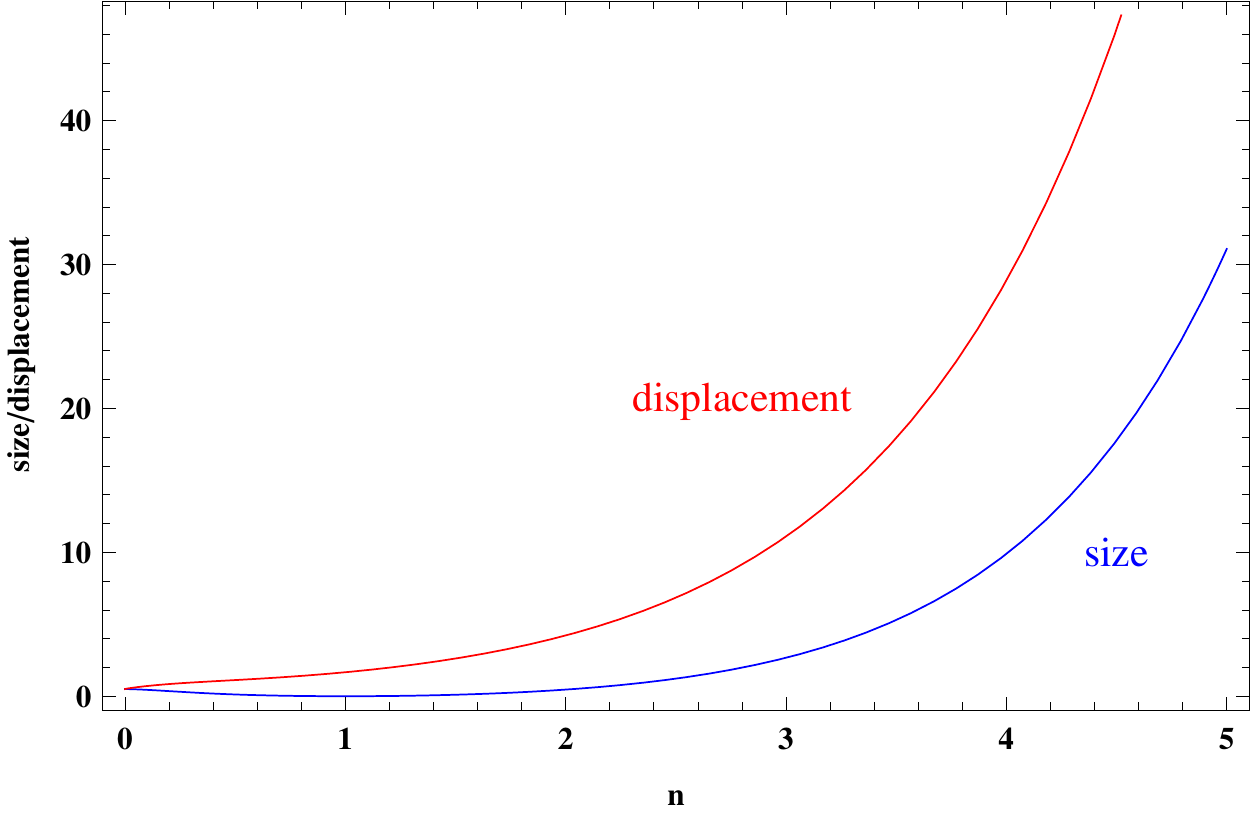}
\caption{Real and imaginary parts of the caustic points calculated in Eq. (\ref{causec}). The imaginary part is a measure of the distance of the secondary caustics, while the real part is a measure of the transverse size. The graphic is for $s=0.6$.}\label{6}
\end{figure}

More in detail, we can make a plot of the real and imaginary parts of Eq. (\ref{causec}) at fixed separation as functions of $n$, as shown in Fig. \ref{6}. This plot makes us appreciate the impressive sizes reached by these giant caustics.

\subsection{Extremely unequal-strength ratio limit}

The limit $\epsilon_B \ll \epsilon_A$ takes the form of the planetary limit in the $n=1$ case \cite{Dom,Boz1,Boz2}. This limit is also interesting because it is possible to fully describe the huge triangular caustics of the close topology, confirming our previous partial results.

\subsubsection{Central caustic}
We consider the main lens at position $ z_{A}=0 $ accompanied by a weak perturbing object placed at $ z_{B}=-s $. The lens equation and the Jacobian determinant take the forms already calculated in the wide limit (\ref{LEwide}), (\ref{Jacwide}). We then introduce the parametrization
\begin{equation}
z=\epsilon_{A}^{1/(n+1)}(1+ \delta) e^{i\theta},
\end{equation}
where $ \delta $ is of order of  $ \epsilon_{B} \ll \epsilon_A $.
Substituting in Eq. (\ref{Jacwide}) and expanding in power series about the point $ \epsilon_{B}=0 $, to first order in $ \epsilon_{B} $, we can solve for $ \delta $:
\begin{equation}
\delta=\dfrac{\epsilon_{B}\lbrace 2 \epsilon_{A}^{2/(n+1)} +4  \epsilon_{A}^{1/(n+1)} \cos \theta  \; s + [1-n+(n+1) \cos (2\theta) ]s^{2}\rbrace}{2(n+1)\big[\epsilon_{A}^{2/(n+1)} +2  \epsilon_{A}^{1/(n+1)} \cos\theta \; s  +s^{2}\big]^{\frac{n+3}{2}}}
\end{equation}
Once determined $ \delta $, we can find the caustic by use of the lens equation:
\begin{equation}
\zeta=-\dfrac{   \lbrace 4\epsilon_{A}^{2/(n+1)}e^{\imath\theta}-\epsilon_{A}^{1/(n+1)}[n-3- 2 (n+3)e^{2\imath\theta}+ (n+1)e^{4\imath\theta}]  \; s+ 4e^{\imath\theta} s^{2}\rbrace \epsilon_{B} \; s}{4 e^{\imath\theta}(\epsilon_{A}^{2/(n+1)} +2  \epsilon_{A}^{1/(n+1)}  \cos \theta \; s  +s^{2})^{\frac{n+3}{2}}} \label{cencau}
\end{equation}
to first order in $ \epsilon_{B} $.

The size of the caustic is proportional to $\epsilon_B$, as in the standard $n=1$ case. More precisely, the distance between the left and right cusps (obtained with $\theta=\pi$ and $\theta=0$ respectively) is
\begin{equation}
\Delta\zeta=\epsilon_Bs\left[\left|s-\epsilon_A^{1/(n+1)}\right|^{-n-1}-\left|s+\epsilon_A^{1/(n+1)}\right|^{-n-1}\right].
\end{equation}
While the size increases with $n$ for $s<2$ because of the first term, it decreases for $s>2$.

It is finally interesting to extend the famous wide/close degeneracy for the central caustic of the ``planetary'' system to the case of arbitrary $n$ \cite{GriSaf,Dom,Boz1}. Indeed, we find that the formula (\ref{cencau}) is invariant under the transformation
\begin{equation}
s\rightarrow \frac{\epsilon_A^{2/(n+1)}}{s}, \;\; \epsilon_B\rightarrow \epsilon_B \left(\frac{\epsilon_A^{1/(n+1)}}{s}\right)^{n-1}.
\end{equation}
We remind that the radius of the critical curve of an isolated exotic object is proportional to $\epsilon_A^{1/(n+1)}$, which makes clear that the inversion of $s$ is exactly the same we find in the standard case. However, for $n\neq 1$, the transformation must be complemented by a change in the strength of the perturbing object.

\subsubsection{Caustics of the perturbing object}

In order to study the caustics of the perturbing object, which reduce to the planetary caustics in the $n=1$ limit \cite{Boz1,Boz2}, we put the main lens at $ z_{A}=-s $ and the secondary lens at $ z_{B}=0$. The lens equation takes the form:
\begin{equation}
\zeta=z-\frac{\epsilon_{A}}{(z+s)^{\frac{n-1}{2}}(\bar{z}+s)^{\frac{n+1}{2}}}-\frac{\epsilon_{B}}{z^{\frac{n-1}{2}}\bar{z}^{\frac{n+1}{2}}},
\end{equation}
and the Jacobian determinant is
\begin{eqnarray}
J &=\left\lbrace 1 + \dfrac{n-1}{2}\left[ \dfrac{\epsilon_{A}}{(z+s)^{\frac{n+1}{2}}(\bar{z}+s) ^{\frac{n+1}{2}}}+\dfrac{\epsilon_{B}}{z^{\frac{n+1}{2}}\bar{z} ^{\frac{n+1}{2}}} \right] \right\rbrace ^{2} \nonumber \\
& \quad -\dfrac{(n+1)^{2}}{4}\left\vert  \dfrac{\epsilon_{A}}{(z+s)^{\frac{n+3}{2}}(\bar{z}+s)^{\frac{n-1}{2}}}+\dfrac{\epsilon_{B}}{z^{\frac{n+3}{2}}\bar{z} ^{\frac{n-1}{2}}} \right\vert^{2}. \label{Jacplan}
\end{eqnarray}

We write
\begin{equation}
z=\rho^{1/(n+1)}\epsilon_{B}^{1/(n+1)}e^{i\theta},
\end{equation}
and substitute in the equation (\ref{Jacplan}). The lowest order Jacobian is:
\begin{equation}
\dfrac{(n+\rho)(\rho-1)}{\rho^{2}}+\dfrac{ [ (n-1) ( 2 \rho+ n -1)-  (n+1)^2 \cos (2 \theta )]\epsilon_{A}}{2\rho s^{n+1}}-\dfrac{n \epsilon_{A}^{2}}{s^{2n+2}}=0.
\end{equation}

Solving this last equation for $ \rho $, we find two solutions:
\begin{eqnarray}
\rho_{\pm} &=& \dfrac{\lbrace \epsilon_{A}[(n+1)^{2}\cos (2\theta)- (n-1)^{2}]+ (2-2n)s^{n+1} \pm(n+1)\sqrt{\Delta}\rbrace s^{n+1} }{4(s^{n+1}-\epsilon_{A})(s^{n+1}+n\epsilon_{A})} \nonumber \\
\Delta &=&  4 s^{2n+2}+ 2 \epsilon_{A}\lbrace(4n-4)s^{n+1}+\epsilon_{A}[n^{2}-6n +1- (n+1)^{2} \cos (2\theta)]\rbrace\sin^{2} \theta. \label{rhoplan}
\end{eqnarray}

Now, when the secondary object is outside of the main lens Einstein ring, the critical curve assumes the shape of an elongated ring. When the planet is inside the main lens critical curve, the secondary object generates two specular ovals. According to the double sign, two branches are present. For external objects ($s>1$) only the higher is real, while for internal objects ($s<1$) both branches are real in a small interval centered on $\theta=\pi/2$ \cite{Boz1,Boz2}. So, we have fully analytical formulae describing the secondary critical curves of the close topology in the extreme strength-ratio limit, something that was not possible to achieve for arbitrary values of the strength ratio.

Through the lens equation, we can find the caustics
\begin{equation}
\zeta=\dfrac{\epsilon_{A}\epsilon_{B}^{1/(n+1)}\rho[n+1 + (n-1)e ^{2 \imath \theta}]- 2 \epsilon_{A}\rho^{n/(n+1)}e^{\imath \theta }s+2\epsilon_{B}^{1/(n+1)}e ^{2 \imath \theta}(\rho-1) s^{n+1}}{2\rho^{n/(n+1)} e^{\imath \theta } s^{n+1}}. \label{cauplan}
\end{equation}
This formula gives a full description of the caustics in the $\epsilon_B\ll \epsilon_A $ limit for all values of $s$. It can be used to obtain general indications of the size and the displacement of the secondary caustics from the main one.

The displacement along the axis joining the two lenses can be obtained as the mid-point $(\zeta(0)+\zeta(\pi))/2$, which lies at the center of the secondary caustic structure. Taking into account that here we have considered the secondary object as the origin of the reference frame, the full displacement from the main object is
\begin{equation}
\zeta_{center} =s - \dfrac{\epsilon_A}{s^{n}},
\end{equation}
which reduces to the classical expression for the position of the planetary caustic for $n=1$ \cite{Dom,Boz1}.

The extension of the caustic in the wide case in the direction parallel to the lens axis is simply given by $(\zeta(0)-\zeta(\pi))$, which reads
\begin{equation}
\Delta\zeta_{\parallel,wide}=2(1+n)\frac{\epsilon_A \epsilon_B^{1/(1+n)}} {s^{n}\left(s^{1+n}-\epsilon_A\right)^{1/(1+n)}}.
\end{equation}
In the vertical direction, orthogonal to the lens axis,  $(\zeta(\pi/2)-\zeta(-\pi/2))$ gives
\begin{equation}
\Delta\zeta_{\perp,wide}=2(1+n)\frac{\epsilon_A \epsilon_B^{1/(1+n)}} {s^{n}\left(s^{1+n}+n\epsilon_A\right)^{1/(1+n)}}.
\end{equation}
The astroid caustic is slightly elongated toward the main lens. By increasing $n$, we have terms in the numerator that become larger and the $s^{n+1}$ in the denominator which tend to compensate and dominate at larger separations, so that the size of the caustic tends to decrease in this regime.

In the close regime $s<\epsilon_A^{1/(1+n)}$, we can easily pick the position of the outer cusp (the most distant one from the lens axis). This is found by setting $\theta=\pi/2$ in Eq. (\ref{cauplan}) and choosing the positive sign for the square root in $\rho$ in Eq. (\ref{rhoplan}). In fact, it can be easily verified that $d\zeta/d\theta=0$ in this point, as it required by the cusp condition. The distance of this cusp from the lens axis is
\begin{equation}
Im[\zeta_+(\pi/2)]=(1+n)\frac{\epsilon_A \epsilon_B^{1/(1+n)}} {s^{n}\left(\epsilon_A-s^{1+n}\right)^{1/(1+n)}}.
\end{equation}

The position of the remaining two cusps cannot be found analytically, but the mid-point is again given by $\zeta(\pi/2)$ with the negative sign for the square root of $\rho$ in Eq. (\ref{rhoplan}):
\begin{equation}
Im[\zeta_-(\pi/2)]=(1+n)\frac{\epsilon_B^{1/(1+n)}} {s^{n}\left[n(\epsilon_A-s^{1+n})\right]^{n/(1+n)}}. \label{transverseclose}
\end{equation}
The difference $Im[\zeta_+(\pi/2)]-Im[\zeta_-(\pi/2)]$ is thus a measure of the transverse size of the triangular secondary caustics.

\begin{figure}[t]
\centering
\includegraphics[width=6cm]{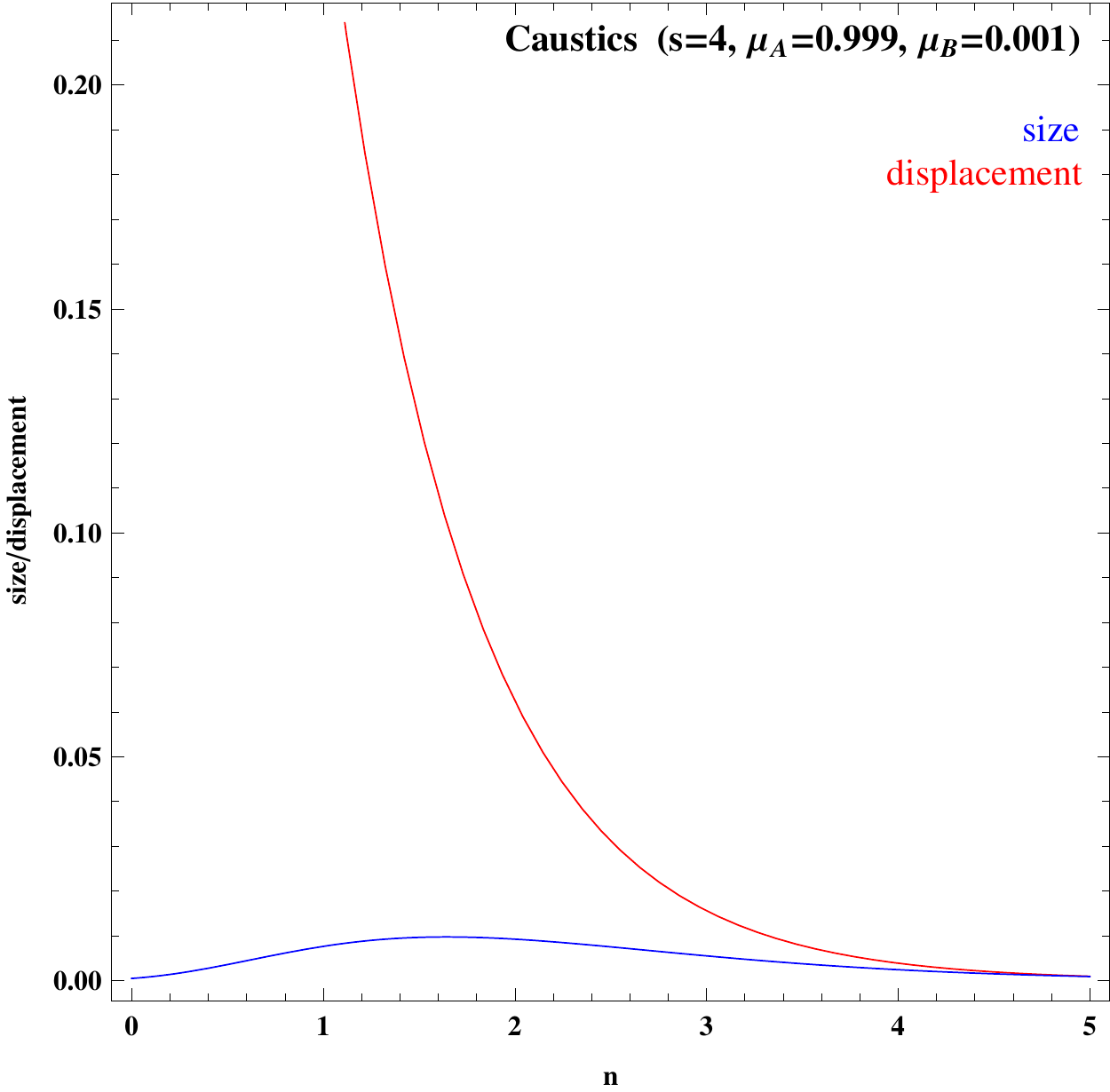}\hspace{4 mm}\includegraphics[width=5.9cm]{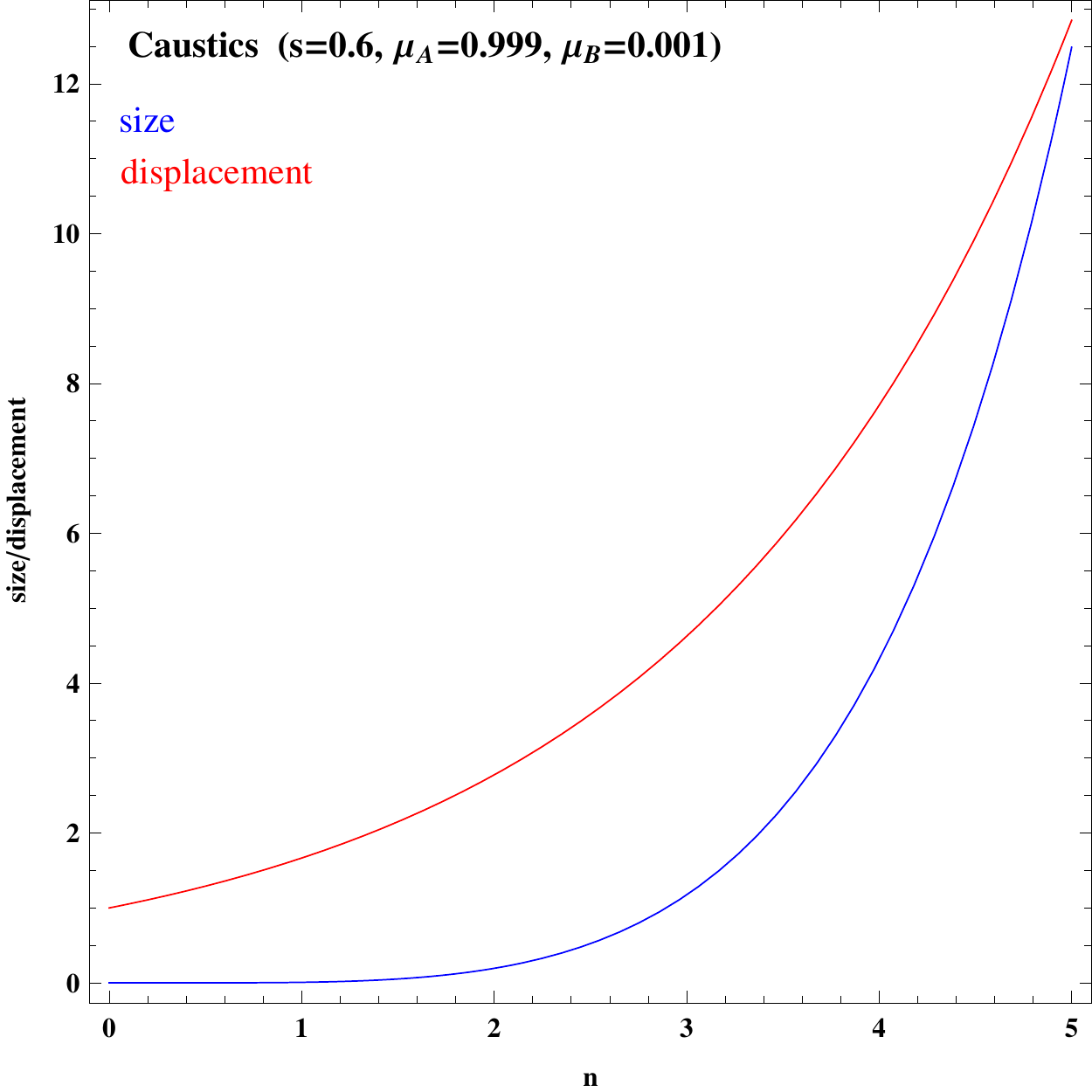}
\caption{Size and displacement of the secondary caustics in the extreme strength ratio limit. } \label{10}
\end{figure}

In Fig. \ref{10} we show the trend of the position and the size of the caustic for $ n \in [0, 5] $.

In the wide regime, we note that the displacement of the secondary caustic from the perturbing object rapidly decreases when going toward higher $n$. The size of the caustic first increases and then decreases at higher $n$.

In the close regime, both the displacement and size increase with $n$. In particular, we see that the size reaches the same order of the displacement, giving rise to the huge caustics we have seen throughout our investigation. For $n<1$, the transverse size (\ref{transverseclose}) shrinks to zero for $s=(1-n)^{1/(n+1)}$, as prescribed by the correct limit of the elliptic umbilic singularity (\ref{Equmb}), and then grows again.

\section{Conclusions}

The mathematical structure of the lens equation has inspired many studies in the past century, since it involves very spectacular aspects of the catastrophe theory \cite{PLW}. Our study of the binary lens made up of two objects with metric falling as $1/r^n$ is not an exception in this picture. We have led a full numerical exploration of all topology regimes in the parameter space $(s,q)$ where $s$ is the separation of the two lenses and $q$ is the strength ratio. We have traced the boundaries of the three topology regimes and for the occurrence of elliptic umbilics, and we have been able to derive analytical approximations for all the extreme cases, in analogy with what has been previously done for the standard binary lens with ordinary gravitational potentials $1/r$ \cite{Boz2,Boz3}.

Although we have not found any new topologies with respect to the standard binary lens, all the positions and sizes of the caustics are modified by the power of the gravitational potential. In particular, the formerly small triangular caustics of the close topology become huge in size for $n>1$, which comes as the most intriguing surprise of our study. Thanks to our analytical approximations, we have been able to quantify the sizes and displacements of caustics in all limits, confirming the numerical study. The by-products of these formulae have been barely explored in this paper. We can just mention the extension of the famous wide/close degeneracy of the planetary limit \cite{GriSaf} to the extreme strength ratio regime of the general $1/r^n$ space.

The applications of this study are two-fold. The regime $n<1$ corresponds to a generic matter distribution with a density profile scaling as $r^{n+2}$. Our exploration provides a considerable extension of previous results only known for the singular isothermal sphere ($n=0$) \cite{ShinEvans}. We have confirmed that an elliptic umbilic catastrophe occurs in the close regime, while no critical curves ever reach the lens position as in the singular isothermal sphere.

On the other side of the Schwarzschild lens, for $n>1$, we have reached our goal of a first exploration of gravitational lensing by exotic $1/r^n$ objects in a non-trivial configuration. Such objects can only be obtained by some violation of the energy conditions. In this paper we have focused on a pair of objects of the same type, so as to understand all modifications of the standard binary case that have to be ascribed by the change of the Schwarzschild potential rather than to mixing effects. A possible next step would be the study of the mixed case with an ordinary Schwarzschild lens around an $1/r^n$ object. Indeed we might expect even more surprises from this asymmetric situation.

The caustic structure is the basis to understand the full phenomenology of gravitational lensing. The much broader extension of caustics in the close and intermediate topology we find in our study suggests that it is much easier to form additional images for binary exotic lenses than for ordinary Schwarzschild lenses. However, this effect has to be weighted by the defocusing power which is known to appear with $1/r^n$ potentials for any $n>1$. So, the application to astrophysical situations might set aside even more interesting surprises. We can imagine that microlensing by such binary lenses would feature incredibly long caustic crossings where they are not expected to be, accompanied by depressed regions in which one or more images are de-magnified. The existence of more signatures that might allow the unambiguous identification of exotic objects, including Ellis wormholes, is indeed the primary goal of this kind of studies. Considering that searches for Ellis wormholes in existing lensing databases have started \cite{TakAsa}, a full understanding of the interaction of these exotic objects with their environment is important to achieve a fully correct interpretation of these data.

\end{document}